\documentclass[9pt]{elsarticle}

\journal{Information and Computation}

\usepackage{amssymb, stmaryrd}

\newcommand {\qc}[1] {{\sf{#1}}}
\def\>{\ensuremath{\rangle}}
\def\<{\ensuremath{\langle}}
\def\h{\ensuremath{\mathcal{H}}}
\def\lh{\ensuremath{\mathcal{L(H)}}}
\def\dh{\ensuremath{\mathcal{D(H}_2)}}

\def\r{\ensuremath{\mathcal{R}}}

\def\a{\ensuremath{\mathcal{A}}}
\def\e{\ensuremath{\mathcal{E}}}

\def\c{\ensuremath{\mathcal{C}}}
\def\d{\ensuremath{\mathcal{D}}}

\renewcommand{\theenumi}{(\arabic{enumi})}

\newcommand {\nil} {\mbox{\bf{nil}}}
\newcommand {\iif} {\mbox{\bf{if}}}
\newcommand {\then} {\mbox{\bf{then}}}

\newcommand{\tr}{{\rm tr}}
\newcommand{\rto}[1]{\stackrel{#1}\rightarrow}
\newcommand{\srto}[1]{\stackrel{#1}\rightarrow_C}
\newcommand{\Rto}[1]{\stackrel{#1}\Rightarrow_C}
\newcommand{\nrto}[1]{\stackrel{#1}\nrightarrow}

\newtheorem{thm}{Theorem}[section]
\newtheorem{cor}{Corollary}[section]
\newtheorem{lem}{Lemma}[section]

\newtheorem{defn}{Definition}[section]

\newtheorem{exmp}{Example}[section]

\begin{document}

\begin{frontmatter}
\title{Probabilistic bisimulations for quantum processes}
\author{Yuan Feng$^1$, Runyao Duan$^1$, Zhengfeng Ji$^2$, and Mingsheng Ying$^1$}
\address{State Key Laboratory of Intelligent Technology and Systems,
Department of Computer Science and Technology, Tsinghua University,
Beijing, 100084, China, \\
State Key Laboratory of Computer Science, Institute of
Software, Chinese Academy of Sciences, Beijing, 100084, China
}
\begin{abstract}
Modeling and reasoning about concurrent quantum systems is very
important for both distributed quantum computing and quantum
protocol verification. As a consequence, a general framework
formally describing communication and concurrency in complex quantum
systems is necessary. For this purpose, we propose a model named
qCCS. It is a natural quantum extension of classical value-passing
CCS which can deal with input and output of quantum states, and
unitary transformations and measurements on quantum systems. The
operational semantics of qCCS is given in terms of probabilistic
labeled transition system. This semantics has many different
features compared with the proposals in the available literature in
order to describe the input and output of quantum systems which are
possibly correlated with other components. Based on this operational
semantics, the notions of strong probabilistic bisimulation and weak
probabilistic bisimulation between quantum processes are introduced.
Furthermore, some properties of these two probabilistic
bisimulations, such as congruence under various combinators, are
examined.
\end{abstract}

\end{frontmatter}

\section{Introduction}

Much attention has been devoted to quantum computation and quantum
information theory (QCQI) in the last two decades since Feynman
\cite{Fe82} proposed the idea that a quantum mechanical system can
be used to perform computation. Benefiting from the possibility of
superposition of different basis states and the linearity of quantum
operations, quantum computing may provide considerable speedup over
its classical analogue \cite{Sh94,Gr96,Gr97}. To provide techniques
of considering computational problems in a conceptual way, rather
than focusing on the details of low-level implementations, some
authors began to study the design and semantics of quantum
programming languages. Knill made the first step by proposing a set
of basic principles for writing quantum pseudo-codes \cite{Kn96},
while the first real quantum programming language, QCL, is due to
\"{O}mer~\cite{Om98,Om03}. A quantum programming language in the
style of Dijkstra's guarded-command language, qGCL, was designed by
Sanders and Zuliani in~\cite{SZ00,Zu01,Zu05b}. They also presented a
probabilistic predicate transformer semantics and a refinement
calculus for their language. A quantum extension of C++ was proposed
by Bettelli et al~\cite{BCS03}, and it was implemented in the form
of a C++ library. The first functional quantum programming language,
QPL, was proposed by Selinger~\cite{Se04} based on the idea of
classical control and quantum data. For detailed surveys on quantum
programming languages and related researches, we refer to
\cite{Se041} or \cite{Ga06}.

The languages presented so far are, however, mostly designed for
sequential quantum computing, where no communication between
physically separated parties is considered. Design and investigation
of languages which can describe quantum concurrent systems and their
communication behaviors have just begun. On the other hand, although
constructing real quantum computers in which quantum programming can
be applied is very difficult, quantum cryptography
\cite{Ek91,BB84,Be92}, which can provide absolute security in
principle even when it has been attacked by a potential quantum
eavesdropper, has been developed so rapidly that quantum
cryptographic systems became commercially available recently
 \cite{PF04}. So, to some extent the need for a language describing
concurrent systems is more urgent than that for sequential
computations in the realm of quantum computation. Furthermore, a
framework of modeling and reasoning about quantum concurrent systems
will provide techniques to prove the properties, such as correctness
and security, of quantum cryptographic protocols, just as we have
noticed in classical world.

The first step of constructing such a general framework of modeling
quantum concurrent systems was made independently by Jorrand and
Lalire \cite{JL04}, and Gay and Nagarajan \cite{GN05}. In
\cite{JL04}, a process algebra for quantum processes was proposed
which can describe both classical and quantum information passing.
Later on, Lalire presented for their language a probabilistic
branching bisimulation which identifies quantum processes associated
with process graphs having the same branching structure
\cite{La05,La06}. In \cite{GN05}, a language called CQP
(Communicating Quantum Processes), which combined the communication
primitives of pi-calculus from \cite{MP92} with primitives for
unitary transformations and measurements, was defined. One
distinctive feature of CQP is a type system which can guarantee the
physical realizability of quantum processes. However, no equivalence
notions between processes were presented there.

The main purpose of this paper is to propose a different model for
quantum concurrent systems. This model, which we call qCCS, is a
quantum extension of classical value-passing CCS \cite{He91,HI93}.
To avoid no-go operations such as quantum cloning in syntactical
level, we explicitly introduce the notion of free quantum variables,
which intuitively denote the quantum systems a process can
reference. When constructing more complicated processes from simpler
ones, this type of variables must be taken into consideration. For
example, if $q$ is one of the free quantum variables of $P$ then the
process $\qc c!q.P$ is invalid because we cannot reference a quantum
system when it has been output. This is in sharp contrast with
classical variables, as classical values can be copied arbitrarily
so that we can use them even after they have been output. As a
consequence, the syntax of qCCS is more complicated than those in
\cite{GN05} and \cite{JL04}. But a type system as introduced in
\cite{GN05} is not necessary in qCCS. Note also that in \cite{JL04},
there was no such mechanism to avoid invalid quantum processes.

In classical process algebra, both call-by-value and call-by-name
strategies can be adopted in the design of semantics. This
flexibility is partially due to the fact that classical information
can be cloned arbitrarily, and so we can talk about classical
information without explicitly referring to the physical carrier of
the information. Quantum information, however, cannot be perfectly
cloned unless it is known. So the only universal way to realize
quantum information transmission is to transfer the physical system
which carries the information. As a consequence, only call-by-name
semantics can be given in quantum process algebra.

To present the operational semantics of qCCS, we introduce the
notion of configuration which is a pair consisting of a quantum
process and an accompanied context instantiating all free quantum
variables of the process. Intuitively, the context describes the
quantum environment in which the process is performed. The
operational semantics of qCCS is then given as a probabilistic
labeled transition system consisting of configurations. There are
some differences between our approach and the previous ones
presented in literature. The first one is that in our semantics,
transitions are from configurations to probability distributions
over configurations, $i.e.$
\[
\rto{}\subseteq Con\times Act\times D(Con)
\]
where $Con$ is the set of configurations and $D(Con)$ is the set of
finite-support distributions on $Con$. Notice that in \cite{JL04}
and \cite{GN05}, probabilistic choice induced by quantum measurement
was resolved in each step. This was achieved by introducing a new
kind of transition $\rightarrow_p$ to represent an evolution which
is caused by an internal action and occurs with probability $p$. In
this paper, however, we do not resolve any probabilistic choice in
intermediate steps but instead keep the probability information all
the time. The motivation for us to make such a design decision is as
follows. First, transitions defined in this way make our operational
semantics much simpler and more CCS-like; second, it gives us a
convenient way to define combined transitions (resp. combined weak
transitions) which are obtained by probabilistically taking
different transitions with the same source configuration and the
same actions (resp. observable actions). That is, the nondeterminism
resulting from the non-probabilistic choice `+' can be resolved in a
probabilistic manner. This is exactly the basis of strong
bisimulation and weak bisimulation defined in this paper. Finally,
by defining transitions in this way, many notions and techniques
introduced in \cite{SL94} and \cite{SL95} for classical
probabilistic processes can be extended to investigate the
properties of probabilistic bisimulations between quantum processes.

The second difference between our approach of semantics and the
previous ones is the ways of dealing with quantum input, quantum
output, and quantum communication. The quantum input rule presented
in \cite{JL04} can only describe the case when the input system is
initially not correlated with the systems the process holds. We
introduce a new inference rule in this paper to deal with the
general case where these systems are correlated. The rule for
quantum output is also refined to keep track of possible correlation
between an output system and the retained systems. As a consequence,
the quantum communication rule in our qCCS has a very simple and
CCS-like form. Note that in \cite{GN05}, no rules for quantum input
and output were introduced because the authors took the viewpoint
that any input action is necessarily accompanied with an output
action (no matter from another process or the environment). However,
we still think it necessary to present rules describing input and
output, since they give us a compositional way to describe quantum
communication between different components.

The main contribution of this paper is a new notion of (strong and
weak) probabilistic bisimulation between quantum processes. As
mentioned above, Lalire \cite{La05} has proposed a notion of
probabilistic branching bisimulation. Our bisimulations, however,
are based on different probabilistic labeled transition system and
motivated by different considerations: First, for two bisimilar
configurations, any action performed by one configuration can be
simulated by a combined action of the other. That is, different
transitions with the same source configuration and the same action
can be chosen simultaneously with different probabilities to
simulate a single transition. Second, the final states of the
quantum contexts when all matching actions have been executed must
be the same when we want to check if two configurations are
bisimilar. We add this requirement because unitary transformations
and measurements are both considered as internal actions, and the
effects of these kinds of actions can be fully reflected only by the
state change of quantum contexts. Finally, note that in qCCS, a
transition from a configuration generally leads to  a finite-support
distribution over configurations, and from each resulted
configuration, different configurations can again be derived with
different probabilities. As a consequence, the execution of a
sequence of actions from a quantum configuration typically forms a
tree rather than a linear path as in classical non-probabilistic
case; any internal actions along any branch of the tree should be
ignored when weak probabilistic bisimulation is concerned.

\subsection{Overview of this paper}

This paper is organized as follows: in Section 2, we review some
basic notions from linear algebra and quantum mechanics which will
be used in this paper. The syntax and operational semantics of qCCS
are presented in Section 3. First, we define inductively quantum
processes and at the same time free quantum variables associated
with each process. Then the notion of configuration is introduced in
which free quantum variables are instantiated by the accompanied
quantum context. The operational semantics of qCCS is given in terms
of probabilistic labeled transition system consisting of
configurations. To show the expressive power of qCCS, we describe
the well-known quantum teleportation protocol with qCCS and show
that it indeed teleports any qubit from one party to another.
Finally, ordinary one-step transitions are extended to combined
multi-step transitions by probabilistically taking different
transitions at each intermediate step.

Section 4 and Section 5 are the main parts of the present paper. We
define the notions of strong and weak probabilistic bisimulations
between configurations and then lift them to bisimulations between
quantum processes. Some properties of these two bisimulations are
also derived. Particularly, we show that probabilistic bisimilarity
is the largest probabilistic bisimulation on $Con$; a weak version
of the congruence property is proved in which bisimilarity of $P$
and $Q$ implies bisimilarity of $P\|R$ and $Q\|R$ for any quantum
process $R$, if either $P$ and $Q$ are free of quantum input or $R$
is free of unitary transformation and quantum measurement. An
example is also presented to show why the standard proof technique
for establishing the preservation of bisimilarity under parallel
combinator in classical CCS cannot be used to prove the result in
general quantum case when the (non-commutative) quantum operations
performed by parallel processes can be interweaved, although it
works well in the two special cases mentioned above.

Section 6 is the concluding section in which we outline the main
results and point out some problems for further study.

\section{Preliminaries}
For convenience of the reader, we briefly recall some basic notions
from linear algebra and quantum theory which are needed in the
sequel. We refer to \cite{NC00} for more details.

\subsection{Basic linear algebra}
A Hilbert space $\h$ is a vector space equipped with an inner
product which in turn is a mapping
$\langle\cdot|\cdot\rangle:\h\times \h\rightarrow \mathbf{C}$
satisfying the following properties:
\begin{enumerate}
\item
$\langle\psi|\psi\rangle\geq 0$ for any $|\psi\>\in\h$, with
equality if and only if $|\psi\rangle =0$;
\item
$\langle\phi|\psi\rangle=\langle\psi|\phi\rangle^{\ast}$;
\item
$\langle\phi|\sum_i\lambda_i|\psi_i\rangle=
\sum_i\lambda_i\langle\phi|\psi_i\rangle$,
\end{enumerate}
where $\mathbf{C}$ is the set of complex numbers, and for each
$\lambda\in \mathbf{C}$, $\lambda^{\ast}$ stands for the complex
conjugate of $\lambda$. For any vector $|\psi\rangle\in\h$, its
length $|||\psi\rangle||$ is defined to be
$\sqrt{\langle\psi|\psi\rangle}$, and it is said to be normalized if
$|||\psi\rangle||=1$. Two vectors $|\psi\>$ and $|\phi\>$ are
orthogonal if $\<\psi|\phi\>=0$. An orthonormal basis of a Hilbert
space $\h$ is a basis $\{|i\rangle\}$ where each $|i\>$ is
normalized and any pair of them are orthogonal.

Let $\lh$ be the set of linear operators on $\h$.  For any $A\in
\lh$, we have the following definitions:
\begin{enumerate}
\item
A non-zero vector $|\psi\>\in \h$ is an eigenvector of $A$ with the
corresponding eigenvalue $\lambda\in \mathbf{C}$ if
$A|\psi\>=\lambda|\psi\>$. We write $spec(A)$ for the set of
eigenvalues of $A$, and call it the spectrum of $A$.
\item
 $A$ is Hermitian if $A^\dag=A$ where
$A^\dag$ is the adjoint operator of $A$ such that
$\<\psi|A^\dag|\phi\>=\<\phi|A|\psi\>^*$ for any
$|\psi\>,|\phi\>\in\h$. The fundamental spectrum theorem states that
the set of all normalized eigenvectors of a Hermitian operator in
$\lh$ contains an orthonormal basis for $\h$. That is, there exists
a so-called spectral decomposition for each Hermitian $A$ such that
$$A=\sum_i\lambda_i |i\>\<i|=\sum_{i\in spec(A)}\lambda_i P_i$$
where the set $\{|i\>\}$ constitute an orthonormal basis of $\h$,
and $P_i=\sum_{j:A|j\>=\lambda_i|j\>}|j\>\<j|$ is the projector to
the corresponding eigenspace of $\lambda_i$.
\item
$A$ is positive if $\<\psi|A|\psi\>\geq 0$ for all $|\psi\>\in\h$;
it is positive-definite if for any nonzero vector $|\psi\>$,
$\<\psi|A|\psi\> > 0$. Note that a positive operator is also
Hermitian.
\item
$A$ is unitary if $A^\dag A=A A^\dag=I_\h$ where $I_\h$ is the
identity operator in $\lh$. In the examples of this paper, we will
use some well-known unitary operators listed as follows: the $CNOT$
operator performed on two qubits such that
$$CNOT=\left(%
\begin{array}{cccc}
  1 & 0 & 0 & 0 \\
  0 & 1 & 0 & 0 \\
  0 & 0 & 0 & 1 \\
  0 & 0 & 1 & 0
\end{array}%
\right),$$ and the 1-qubit Hadamard operator $H$ and Pauli operators
$\sigma_0,\sigma_1,\sigma_2,\sigma_3$ defined respectively as
\[
H=\frac{1}{\sqrt{2}}\left(%
\begin{array}{cc}
  1 & 1 \\
  1 & -1 \\
\end{array}%
\right),\ \  \sigma_0=I=\left(%
\begin{array}{cc}
  1 & 0 \\
  0 & 1 \\
\end{array}%
\right),
\]

\[
\sigma_1=\left(%
\begin{array}{cc}
  0 & 1 \\
  1 & 0 \\
\end{array}%
\right),\ \sigma_2=\left(%
\begin{array}{cc}
  0 & -i \\
  i & 0 \\
\end{array}%
\right),\ \sigma_3=\left(%
\begin{array}{cc}
  1 & 0 \\
  0 & -1 \\
\end{array}%
\right).
\]

\item
The trace of $A$ is defined as $\tr(A)=\sum_i \<i|A|i\>$ for some
given orthonormal basis $\{|i\>\}$ of $\h$. It is worth noting that
trace function is actually independent of the orthonormal basis
selected. It is also easy to check that trace function is linear and
$\tr(AB)=\tr(BA)$ for any operators $A,B\in \lh$.
\end{enumerate}

Let $\h_1$ and $\h_2$ be two Hilbert spaces of dimensions $n_1$ and
$n_2$, respectively. Then their tensor product $\h_1\otimes \h_2$ is
defined as an $n_1 n_2$-dimensional vector space consisting of
linear combinations of the vectors
$|\psi_1\psi_2\rangle=|\psi_1\>|\psi_2\rangle =|\psi_1\>\otimes
|\psi_2\>$ with $|\psi_1\rangle\in \h_1$ and $|\psi_2\rangle\in
\h_2$. Here the tensor product of two vectors is defined by a new
vector such that
$$\left(\sum_i \lambda_i |\psi_i\>\right)\otimes
\left(\sum_j\mu_j|\phi_j\>\right)=\sum_{i,j} \lambda_i\mu_j
|\psi_i\>\otimes |\phi_j\>.$$ Then $\h_1\otimes \h_2$ is also a
Hilbert space where the inner product is defined as the following:
for any $|\psi_1\>,|\phi_1\>\in\h_1$ and $|\psi_2\>,|\phi_2\>\in
\h_2$,
$$\<\psi_1\otimes \psi_2|\phi_1\otimes\phi_2\>=\<\psi_1|\phi_1\>_{\h_1}\<
\psi_2|\phi_2\>_{\h_2}$$ where $\<\cdot|\cdot\>_{\h_i}$ is the inner
product of $\h_i$. For any $A_1\in \mathcal{L}(\h_1)$ and $A_2\in
\mathcal{L}(\h_2)$, $A_1\otimes A_2$ is defined as a linear operator
in $\mathcal{L}(\h_1 \otimes \h_2)$ such that for each
$|\psi_1\rangle \in \h_1$ and $|\psi_2\rangle \in \h_2$,
$$(A_1\otimes A_2)|\psi_1\psi_2\rangle = A_1|\psi_1\rangle\otimes
A_2|\psi_2\rangle.$$  The partial trace of $A\in\mathcal{L}(\h_1
\otimes \h_2)$ with respected to $\h_1$ is defined as
$\tr_{\h_1}(A)=\sum_i \<i|A|i\>$ where $\{|i\>\}$ is an orthonormal
basis of $\h_1$. Similarly, we can define the partial trace of $A$
with respected to $\h_2$. Partial trace functions are also
independent of the orthonormal basis selected.

A linear operator $\e$ on $\lh$ is completely positive if it maps
positive operators in $\mathcal{L}(\h)$ to positive operators in
$\mathcal{L}(\h)$, and for any auxiliary Hilbert space $\h'$, the
trivially extended operator $\mathcal{I}_{\h'}\otimes \e$ also maps
positive operators in $\mathcal{L(H'\otimes H)}$ to positive
operators in $\mathcal{L(H'\otimes H)}$. Here $\mathcal{I}_{\h'}$ is
the identity operator on $\mathcal{L(H')}$. The elegant and powerful
Kraus representation theorem \cite{Kr83} of completely positive
operators states that a linear operator $\e$ is completely positive
if and only if there are some set of operators $\{E_i,
i=1,\dots,d\}$ with appropriate dimension such that
$$
\e(A)=\sum_{i=1}^d E_iA E_i^\dag
$$
for any $A\in \lh$. The operators $E_i$ are called Kraus operators
of $\e$. A linear operator is said to be a super-operator if it is
completely positive and trace-preserving. Here an operator $\e$ is
trace-preserving if $\tr(\e(A))= \tr(A)$ for any linear operator
$A$. Then a super-operator is just a completely positive operator
with its Kraus operators $E_i$ satisfying $\sum_i E_i^\dag E_i= I$.

\subsection{Basic quantum mechanics}

According to von Neumann's formalism of quantum mechanics
\cite{vN55}, an isolated physical system is associated with a
(finite-dimensional) Hilbert space which is called the state space
of the system. A pure state of a quantum system is a normalized
vector in its state space, and a mixed state is represented by a
density operator. Here a density operator $\rho$ on Hilbert space
$\h$ is a positive linear operator such that $\tr(\rho)= 1$. Another
equivalent representation of density operator is probabilistic
ensemble of pure states. In particular, given an ensemble
$\{(p_i,|\psi_i\rangle)\}$ where $p_i \geq 0$, $\sum_{i}p_i=1$, and
$|\psi_i\rangle$ are pure states,
$\rho=\sum_{i}p_i|\psi_i\rangle\langle\psi_i|$ is a density
operator. Conversely, each density operator can be generated by an
ensemble of pure states in this way. In this paper, we denote by
$\d(\h)$ the set of density operators on Hilbert space $\h$.

The evolution of a closed quantum system is described by a unitary
operator on its state space: if the states of the system at times
$t_1$ and $t_2$ are $\rho_1$ and $\rho_2$, respectively, then
$\rho_2=U\rho_1U^{\dag}$ for some unitary operator $U$ which depends
only on $t_1$ and $t_2$. In particular, if $\rho_1$ and $\rho_2$ are
pure states $|\psi_1\rangle$ and $|\psi_2\rangle$, respectively,
then we have $|\psi_2\rangle =U|\psi_1\rangle$.

Observation of a quantum system is a quantum measurement represented
by a Hermitian operator $M$ on the associated state space. Suppose
$M$ has the spectral decomposition $M=\sum_m mP_m,$ where $P_m$ is
the projector onto the eigenspace of $M$ associated with eigenvalue
$m$. Then the probability of obtaining measurement result $m$ when
the system is initially in the state $\rho$ is $p_m=\tr(P_m\rho)$,
and if $p_m>0$ then the post-measurement state of the system given
the outcome $m$ becomes
$$\frac{P_m\rho P_m}{p_m}.$$
For the case that $\rho$ is a pure state $|\psi\rangle$, we have
$p_m=\<\psi|P_m|\psi\>$, and the post-measurement state is
$P_m|\psi\>/\sqrt{p_m}$.

The state space of a composite system (for example, a quantum system
consisting of many qubits) is the tensor product of the state spaces
of its components. For a mixed state $\rho$ on $\h_1 \otimes \h_2$,
partial traces of $\rho$ have explicit physical meanings: the
density operators $\tr_{\h_1}\rho$ and $\tr_{\h_2}\rho$ are exactly
the reduced quantum states of $\rho$ on the second and the first
component system, respectively. Note that in general, the state of a
composite system cannot be decomposed into tensor product of the
reduced states on its component systems. A well-known example is the
so-called EPR state
$$\frac{1}{\sqrt{2}}(|00\>+|11\>)
$$
in 2-qubit system. This kind of states is called entangled states.
To see the weirdness of entanglement, suppose a measurement $M=
\lambda_0|0\>\<0|+\lambda_1|1\>\<1|$ is applied on the first qubit
of the EPR state. Then after the measurement, the second qubit will
definitely collapse into state $|0\>$ or $|1\>$ depending on whether
the outcome $\lambda_0$ or $\lambda_1$ is observed. In other words,
the measurement on the first qubit changes the state of the second
qubit in a way. This is an outstanding feature of quantum mechanics
which has no counterpart in classical world, and is the key to many
quantum information processing tasks  such as teleportation
\cite{BB93} and superdense coding \cite{BW92}.

\subsection{Quantum no-cloning theorem}

Classical information can be arbitrarily cloned. However, the
linearity of quantum operations prohibits the possibility of
perfectly cloning an unknown quantum state \cite{WZ82}. The formal
argument goes as follows. Suppose a quantum cloning device is
possible, $i.e.$ there is a physically realizable procedure such
that the transformation
\begin{equation}\label{eq:cloning}
|\psi\>|\Sigma\>\longrightarrow |\psi\>|\psi\>
\end{equation}
holds for any $|\psi\>\in \h$. Here $|\Sigma\>$ is a standard state
which is independent of $|\psi\>$. In particular, for two orthogonal
states $|0\>$ and $|1\>$, we have
\[
|0\>|\Sigma\>\longrightarrow |0\>|0\>\hspace{2em} \mbox{ and
}\hspace{2em} |1\>|\Sigma\>\longrightarrow |1\>|1\>.\] Now let
$|\psi\>=\alpha|0\>+\beta|1\>$. Because of the linearity of quantum
operations imposed by basic principles of quantum mechanics, we have
\begin{equation}\label{eq:cloning1}
|\psi\>|\Sigma\>=\alpha|0\>|\Sigma\>+\beta|1\>|\Sigma\>
\longrightarrow \alpha|0\>|0\>+\beta|1\>|1\>.
\end{equation}
On the other hand, Eq.(\ref{eq:cloning}) can be rewritten as
\begin{equation}\label{eq:cloning2}
|\psi\>|\Sigma\> \longrightarrow
\alpha^2|0\>|0\>+\beta^2|1\>|1\>+\alpha\beta(|0\>|1\>+|1\>|0\>).
\end{equation}
Comparing the right-hand sides of Eq.(\ref{eq:cloning1}) and
Eq.(\ref{eq:cloning2}), we deduce that $\alpha=0$ or $\beta=0$. That
is, the universal cloning procedure presented in
Eq.(\ref{eq:cloning}) does not exist. This is the well-known quantum
no-cloning theorem.

Quantum no-cloning theorem has been shown to be connected with some
other no-go principles such as no-signaling principle which states
that signals can not be sent faster than the speed of light
\cite{Bu87,GW83}. No-cloning theorem was also used to argue for the
security of quantum cryptography \cite{BB84}. In the scenario of
communication, because unknown quantum states can not be perfectly
cloned, transferring of quantum datum must be done by sending the
physical system which carries the information, unless the datum to
be transmitted is already known to the sender. This is in sharp
contrast with the case in classical world where to send an unknown
datum, one need only produce a copy of it and then transmit the
copy. The sender needs not know the classical datum since perfect
cloning is always possible.

\section{Basic Definitions of qCCS}

In this section, we give the basic definitions of qCCS. Subsections
3.1 and 3.2 are devoted to the syntax and the operational semantics,
respectively. In subsection 3.3, we extend ordinary one-step
transitions to combined multi-step transitions.
\subsection{Syntax}
For the sake of simplicity, we consider only two types of data: the
set of real numbers \qc {Real} for classical data, and the set of
qubits \qc {Qbt} for quantum data. We denote by $cVar$ (ranged over
by $x,y,\dots$) and $qVar$ (ranged over by $q,r,\dots$) the set of
classical variables on \qc {Real} and quantum variables on \qc
{Qbt}, respectively. The set of expressions with the value domain
\qc {Real} is denoted by $Exp$ and ranged over by $e$. Let $cChan$
be the set of classical channel names, ranged over by $c,d,\dots$,
and $qChan$ the set of quantum channel names, ranged over by $\qc
c,\qc d,\dots$. Let $Chan=cChan\cup qChan$. A relabeling function
$f$ is a one to one function from $Chan$ to $Chan$ such that
$f(cChan)\subseteq cChan$ and $f(qChan)\subseteq qChan$.

From these notations, we now propose the syntax of qCCS as follows.
For simplicity, we often abbreviate the indexed set
$\{q_1,\dots,q_n\}$ to $\bar{q}$ when $q_1, \dots,q_n$ are distinct
quantum variables and the dimension $n$ is understood.

\begin{defn}(quantum process)\label{def:qProc}
\rm The set of quantum processes $qProc$ and the free quantum
variable function $qv: qProc\rto{} 2^{qVar}$ are defined inductively
by the following formation rules:
\begin{enumerate}

\item  $\nil \in qProc$, and $qv(\nil)=\emptyset$;

\item  $c?x.P \in qProc$, and $qv(c?x.P)=qv(P)$;

\item  $c!e.P \in qProc$, and $qv(c!e.P)=qv(P)$;

\item  $\qc c?q.P \in qProc$, and $qv(\qc c?q.P)=qv(P)-\{q\}$;

\item  If $q\not \in qv(P)$ then $\qc c!q.P \in qProc$, and $qv(\qc
c!q.P)=qv(P)\cup\{q\}$;

\item  $U[\bar{q}].P\in qProc$, and
$qv(U[\bar{q}].P)=qv(P)\cup\bar{q}$;

\item  $M[\bar{q};x].P\in qProc$, and
$qv(M[\bar{q};x].P)=qv(P)\cup\bar{q}$;

\item  $P+Q\in qProc$, and $qv(P+Q)=qv(P)\cup qv(Q)$;

\item  If $qv(P)\cap qv(Q)=\emptyset$ then $P\| Q\in qProc$, and
$qv(P\| Q)=qv(P)\cup qv(Q)$;

\item  $P[f]\in qProc$, and $qv(P[f])=qv(P)$;

\item  $P\backslash L\in qProc$, and $qv(P\backslash L)=qv(P)$;

\item $\iif\ b\ \then\ P\in qProc$, and $qv(\iif\ b\ \then\
P)=qv(P)$,

\end{enumerate}
where $P,Q\in qProc$, $c\in cChan$, $x,y\in cVar$, $\qc c\in qChan$,
$q,q_1,\dots,q_n\in qVar$, $e\in Exp$, $f$ is a relabeling function,
$L\subseteq Chan$, $b$ is a boolean-valued expression, $U$ is a
unitary operator, and $M$ is a Hermitian operator.

\end{defn}

The process constructs we give here are quite similar to those in
classical CCS, and they also have similar intuitive meanings: $\nil$
stands for a process which does not perform any action; $c?x$ and $
c!e$ are respectively classical input and classical output, while
$\qc c?q$ and $\qc c!q$ are their quantum counterparts. $U[\bar{q}]$
denotes the action of performing a unitary transformation $U$ on the
qubits $\bar{q}$ while $M[\bar{q};x]$ measures the qubits $\bar{q}$
according to $M$ and stores the measurement outcome into the
classical variable $x$. $+$ models nondeterministic choice: $P+Q$
behaves like either $P$ or $Q$ depending on the choice of the
environment. $\|$ denotes the usual parallel composition. The
operators $\backslash L$ and $[f]$ model restriction and relabeling,
respectively: $P\backslash L$ behaves like $P$ as long as any action
through the channels in $L$ is forbidden, and $P[f]$ behaves like
$P$ where each channel name is replaced by its image under the
relabeling function $f$. Finally, $\iif\ b\ \then\ P$ is the
standard conditional choice where $P$ can be executed only if $b$ is
true.

For any quantum process $P$, $qv(P)$ is exactly the set of quantum
variables which $P$ can reference. Note that in the process
 $\qc c!q.P$, the assumption $q\not\in qv(P)$ guarantees
that a quantum system will not be referenced after it has been
output. This is a requirement of quantum no-cloning theorem. For the
same reason, we assume $q_1,\dots,q_n$ distinct in $U[\bar{q}].P$
and $M[\bar{q};x].P$ (Recall that the notation $\bar{q}$ implies
that $q_1,\dots,q_n$ are distinct). Furthermore, since we intend to
use parallel combinator $\|$ to model separate parties which can
perform actions locally on their own systems and communicate with
each other through channels, the assumption $qv(P)\cap
qv(Q)=\emptyset$ guarantees that $P$ and $Q$ will never reference a
quantum system simultaneously.

The notion of free classical variables in quantum processes can be
defined in the usual way with a unique modification that quantum
measurement $M[\bar{q};x]$ has binding power on $x$. A quantum
process $P$ is closed if it contains no free classical variables,
$i.e.$, $fv(P)=\emptyset$.

\subsection{Operational semantics of qCCS}

To present the operational semantics of qCCS, we first introduce the
notion of configuration. Note that for any $P\in qProc$ with
$fv(P)\subseteq \{x_1,\dots,x_n\}$ and any indexed set
$\bar{v}=\{v_1,\dots,v_n\}$ of real values, the process
$P[\bar{v}/\bar{x}]$ obtained by instantiating classical variables
$\bar{x}$ with $\bar{v}$ is closed. The following definition
introduces a corresponding instantiation for free quantum variables.
Similar notions were also presented in \cite{JL04} and \cite{GN05}
in a somewhat different way.

\begin{defn}(Configuration) For any closed quantum process $P$, if
$qv(P)\subseteq \bar{q}$ then a pair of the form
\begin{equation}\label{def:configuration}
<P;\bar{q}=\rho>
\end{equation}
is called a configuration, where $\rho$ is a density operator in
$2^n$-dimensional Hilbert space and $n$ is the length of $\bar{q}$.
The set of configurations is denoted by $Con$ and ranged over by
$\c,\d,\dots$. In the configuration $\c=<P;\bar{q}=\rho>$,
`$\bar{q}=\rho$' is called the quantum context of $\c$ and denoted
$Context(\c)$.
\end{defn}

Intuitively, quantum context describes the `quantum environment' in
which a process lives. All of the quantum systems which a process
can reference must be included in the accompanied quantum context.

Let $D(Con)$ be the set of finite-support probability distributions
over $Con$, $i.e.$ $$ D(Con)=\{\mu:Con\rightarrow [0,1]\ |\
\mu(\c)>0 \mbox{ for finitely many $\c$, and
}\sum_{\mu(\c)>0}\mu(\c)= 1\}.
$$
For any $\mu\in D(Con)$, we denote by $supp(\mu)$ the support set of
$\mu$, $i.e.$ the set of configurations $\c$ such that $\mu(\c)>0$.
When $\mu$ is a simple distribution such that $supp(\mu)=\{\c\}$ for
some $\c$, we abuse the notation slightly to denote $\mu$ by $\c$.
Just as in \cite{JL04} and \cite{GN05}, sometimes we find it
convenient to denote a distribution $\mu\in D(Con)$ by an explicit
form $\mu= \boxplus_{i\in I}p_i\bullet \c_i$ (or $\mu= \boxplus
p_i\bullet \c_i$ when the index set $I$ is understood) where
$supp(\mu)=\{\c_i\ |\ i\in I\}$ and $\mu(\c_i)=p_i$ for each $i\in
I$. Given $\mu_1,\dots,\mu_n\in D(Con)$ and $p_1,\dots,p_n\in
(0,1]$, $\sum_i p_i=1$, we define the combined distribution, denoted
by $\sum_{i=1}^n p_i\mu_i$, to be a new distribution $\mu\in D(Con)$
such that for any $\d\in supp(\mu)$, $\mu(\d)=\sum_i p_i \mu_i(\d)$.
It is obvious that $supp(\sum_i p_i \mu_i) = \bigcup_i supp(\mu_i)$.

As usual, the operational semantics of qCCS is given in terms of
probabilistic labeled transition system. Let
\begin{eqnarray*}
Act&=&\{c?v,c!v\ |\ c\in cChan, v\in \qc{Real}\}\\
&&\ \cup\ \{\qc c?r,\qc c?r:\rho,\qc c!r\ |\ \qc c\in qChan, r\in qVar, \rho\in \dh\}\ \cup\ \{\tau\}
\end{eqnarray*}
where $\tau$ is the silent action, and $\dh$ is the set of density
operators on a 2-dimensional Hilbert space. Then the semantics of
qCCS is given by the probabilistic labeled transition system
$(Con,Act,\rightarrow)$, where $\rightarrow\subseteq Con\times
Act\times D(Con)$ is the smallest relation satisfying the rules
defined in Definitions 3.3 through 3.13. (For brevity, we write $\c
\rto{\alpha} \mu$ instead of $(\c,\alpha,\mu)\in \rightarrow$).

\begin{defn}\rm (Classical rules)
\[
\begin{array}{rl}
\mbox{\textbf{C-Inp}}: & \frac{}{\displaystyle <c?x.P;C> \rto{c?v}
{<P[v/x];C>}}\hspace{1em} \mbox{for all }v\in \qc {Real}\\
\\
\mbox{\textbf{C-Outp}}: & \frac{}{\displaystyle <c!e.P;C> \rto{c!v}
{<P;C>}} \hspace{1em} \mbox{where $
v$ is the value of $e$}\\
\\
\mbox{\textbf{C-Com}}: & \frac{\displaystyle <P_1;C>\rto{c?v}
{<P_1';C>},\hspace{1em} <P_2;C>\rto{c!v} {<P_2';C>}} {\displaystyle
<P_1\|P_2;C>\rto{\tau}
{<P_1'\|P_2';C>}}\\
\\
 & \frac{\displaystyle <P_1;C>\rto{c!v}
{<P_1';C>},\hspace{1em} <P_2;C>\rto{c?v} {<P_2';C>}} {\displaystyle
<P_1\|P_2;C>\rto{\tau} {<P_1'\|P_2';C>}}
\end{array}
\]
\end{defn}

These three rules describe the passing of classical messages; they
are almost the same as in classical value-passing CCS. Contexts
remain untouched in these rules since they include only the
accompanied $quantum$ systems, which will not be changed by
$classical$ input and output. Other classical rules are incorporated
into Definitions 3.9 through 3.13 below.

\begin{defn}\rm (Quantum-input rules)
\[
\begin{array}{rl}
\mbox{\textbf{Q-Inp1}}: & \frac{}{\displaystyle <\qc
c?q.P;\bar{q}=\rho>
\rto{\qc c?r:\sigma} <P[r/q];r,\bar{q}=\sigma\otimes\rho>} \mbox{where  } r\not\in \bar{q} \mbox{  and  } \sigma\in\dh\\
\\
\mbox{\textbf{Q-Inp2}}: & \frac{}{\displaystyle <\qc
c?q.P;\bar{q}=\rho>
\rto{\qc c?r} {<P[r/q];\bar{q}=\rho>}}\hspace{1.7em} \mbox{where }r\in \bar{q}-qv(\qc c?q.P) \\
\end{array}
\]
\end{defn}

In \cite{JL04}, only a rule similar to the first one was presented
for quantum input. This rule makes sense when the input system
(denoted by the quantum variable $r$) is initially not correlated
(neither entangled nor classically correlated) with the quantum
systems in $\bar{q}$. However, one of the essential features which
distinguish quantum mechanics from classical mechanics is that
different systems can lie in an entangled state which can not be
determined by the reduced states of individual systems. This
argument leads naturally to the following inference rule:
\[
<\qc c?q.P;\bar{q}=\rho> \rto{\qc c?r:\rho'}
{<P[r/q];r,\bar{q}=\sigma>}\hspace{1em} \mbox{where }r\not\in
\bar{q},\ \tr_{\bar{q}}\sigma=\rho',\ \mbox{and }
\tr_{r}\sigma=\rho.
\]
Any quantum input can be characterized by this rule since no
constraints are made on the new state $\sigma$ except
$\tr_{r}\sigma=\rho$ which means that the state of initial systems
remains untouched. This rule is, however, also problematic. First,
it is not image-finite in the sense that from the source
configuration $<\qc c?q.P;\bar{q}=\rho>$ and the action $\qc
c?r:\rho'$, there are infinitely many derived configurations which
satisfy the rule. Second, in general the effect of this transition
on the accompanied context is not a super-operator independent of
$\rho$. This will make some proofs in Sections 4 and 5 infeasible.

In consideration of the above arguments, we present rules
\textbf{Q-Inp1} and \textbf{Q-Inp2} which describe the input of  a
qubit from the outside and the inside of the context, respectively.
Note that the context is kept untouched in rule \textbf{Q-Inp2}. The
intuition behind is that when the system to be input has already
been described in the context, the input action is merely a
declaration that the process can reference this system, which of
course does not change the state of the whole system.

\begin{defn}\rm \textbf{Q-Outp} (Quantum-output rule)
\[
\frac{}{\displaystyle <\qc c!q.P;\bar{q}=\rho> \rto{\qc c!q}
{<P;\bar{q}= \rho>}}\]
\end{defn}

The quantum output rule presented in \cite{JL04} was of the
following form (rewritten with our notations):
\[
{\displaystyle <\qc c!q.P;\bar{q}=\rho> \rto{\qc c!q}
{<P;\bar{q}-\{q\}=\tr_{q} \rho>}}
\]
with the intuition that we do not care about the state of a quantum
system when it has been output. The information about how the output
system is correlated with the systems remained in the context is,
however, totally lost; problems will arise if we input again the
system which was just output. The \textbf{Q-Outp} rule presented
above can deal with this problem since the quantum context remains
unchanged so that any information is kept.

\begin{defn}\rm \textbf{Unit} (Unitary transformation rule)
\[
\frac{}{\displaystyle <U[\bar{r}].P ;\bar{q}=\rho> \rto{\tau}
{<P;\bar{q}=U_{\bar{r}}\rho U_{\bar{r}}^\dag>}}
\]
where $U_{\bar{r}}\rho U_{\bar{r}}^\dag$ denotes the application of
unitary transformation $U$ on the system consisting of $\bar{r}$. To
be specific, let $length(\bar{r})=k$ and $length(\bar{q})=n$. Then
$U_{\bar{r}}=\Pi_{\bar{r}}^\dag (U\otimes I^{\otimes(n-k)})
\Pi_{\bar{r}}$ where $\Pi_{\bar{r}}$ is a permutation which places
$r_1,\dots,r_k$ at the head of $\bar{q}$, and $I$ is the identity
transformation. Similar notations were also introduced in
\cite{JL04}.
\end{defn}

In our framework of qCCS, performing a unitary transformation is
modeled by a $\tau$-action which is unobservable from outside. The
same treatment is applied to measurement on quantum systems.

\begin{defn}\rm \textbf{Meas} (Measurement rule)
\[
\frac{}{\displaystyle <M[\bar{r};x].P ;\bar{q}=\rho> \rto{\tau}
 \boxplus_{i\in I} p_i\bullet<P[\lambda_i/x];\bar{q}=P_{i,\bar{r}}\rho P_{i,\bar{r}}/p_i>}
\]
where $M$ is a Hermitian operator with the spectral decomposition
$M=\sum_{i\in I} \lambda_i P_i$, $P_{i,\bar{r}}$ denotes the
projection $P_i$ performed on the system consisting of $\bar{r}$,
$i.e.$, $P_{i,\bar{r}}=\Pi_{\bar{r}}^\dag (P_i\otimes
I^{\otimes(n-k)}) \Pi_{\bar{r}}$, and $p_i=\tr(P_{i,\bar{r}}\rho)$.
\end{defn}

\begin{defn}\rm \textbf{Q-Com} (Quantum-communication rule)
\[
\frac{\displaystyle <P_1;C>\rto{\qc c?r} {<P_1';C>},\hspace{1em}
<P_2;C>\rto{\qc c!r} {<P_2';C>}}{\displaystyle
<P_1\|P_2;C>\rto{\tau} {<P_1'\|P_2';C>}}
\]
\[
\frac{\displaystyle <P_1;C>\rto{\qc c!r} {<P_1';C>},\hspace{1em}
<P_2;C>\rto{\qc c?r} {<P_2';C>}}{\displaystyle
<P_1\|P_2;C>\rto{\tau} {<P_1'\|P_2';C>}}
\]
\end{defn}

It may be surprising at first glance that there is no communication
rule in which the participating action of either parallel process is
of the form $\qc c?r:\rho$. In other words, quantum input from
outside the accompanied context cannot lead to quantum
communication. The reason is as follows. To make $<P_1\|P_2;C>$ a
valid configuration, the context $C$ must involve all the free
quantum variables occur in $P_1$ and $P_2$. As a consequence, any
qubit which will be input by $P_1$ or $P_2$ during the quantum
communication between them is from the context $C$.

\begin{defn}\rm (Interleaving rules)
\[
\begin{array}{rl}
\mbox{\textbf{Inp-Int}}: & \frac{\displaystyle <P_1;C>\rto{\qc c?r}
<P_1';C'>}{\displaystyle <P_1\|P_2;C>\rto{\qc c?r}
<P_1'\|P_2;C'>}\hspace{1.7em} \mbox{where }r\not\in qv(P_2)
\\ \\
& \frac{\displaystyle <P_2;C>\rto{\qc c?r} <P_2';C'>}{\displaystyle
<P_1\|P_2;C>\rto{\qc c?r} <P_1\|P_2';C'>}\hspace{1.7em} \mbox{where
}r\not\in qv(P_1)
\\ \\ \\
\mbox{\textbf{Oth-Int}}: & \frac{\displaystyle <P_1;C>\rto{\alpha}
\boxplus p_i\bullet <P_1^i;C_i>}{\displaystyle
<P_1\|P_2;C>\rto{\alpha} \boxplus p_i\bullet <P_1^i\|P_2;C_i>}
\hspace{1.7em}\mbox{where $\alpha$ is not of the form } \qc c?r
\\
\\
&\frac{\displaystyle <P_2;C>\rto{\alpha} \boxplus p_i\bullet
<P_2^i;C_i>}{\displaystyle <P_1\|P_2;C>\rto{\alpha} \boxplus
p_i\bullet <P_1\|P_2^i;C_i>}\hspace{1.7em}\mbox{where $\alpha$ is
not of the form } \qc c?r
\end{array}
\]
\end{defn}

The side conditions $r\not\in qv(P_2)$ and $r\not\in qv(P_1)$ in
\textbf{Inp-Int} rules are presented to exclude the possibility that
one process inputs a qubit which is referencing by another parallel
process. Other interleaving rules, including those dealing with
quantum output and classical actions, are incorporated into
\textbf{Oth-Int} rules.

 \vspace{1em} The following
rules are similar to their classical counterparts.

\begin{defn}\rm \textbf{Sum} (Summation rule)
\[
\frac{\displaystyle <P;C>\rto{\alpha} \mu}{\displaystyle
<P+Q;C>\rto{\alpha} \mu}, \hspace{3em} \frac{\displaystyle
<Q;C>\rto{\alpha} \mu}{\displaystyle <P+Q;C>\rto{\alpha} \mu} \\
\]
\end{defn}

\begin{defn}\rm \textbf{Rel} (Relabeling rule)
\[
\frac{\displaystyle <P;C>\rto{\alpha} \boxplus p_i\bullet
<P_i;C_i>}{\displaystyle <P[f];C>\rto{\alpha[f]}  \boxplus
p_i\bullet <P_i[f];C_i>}
\]
\end{defn}

Here we extend the definition of relabeling function to actions and
quantum processes in an obvious way.

\begin{defn}\rm \textbf{Res} (Restriction rule)
\[
 \frac{\displaystyle <P;C>\rto{\alpha}
 \boxplus p_i\bullet
<P_i;C_i>}{\displaystyle <P\backslash L;C>\rto{\alpha}  \boxplus
p_i\bullet <P_i\backslash L;C_i>} \mbox{ where }cn(\alpha)\not\in L
\]
\end{defn}

Here the function $cn$ returns the channel name used by an action.

\begin{defn}\rm \textbf{Cho} (Choice rule)
\[
\frac{\displaystyle <P;C>\rto{\alpha} \mu}{\displaystyle <\iif\ b\
\then\ P;C>\rto{\alpha} \mu} \mbox{ where $b$ is true}
\]
\end{defn}

When $b$ is false then the configuration $<\iif\ b\ \then\ P;C>$
cannot perform any action.

\vspace{1em} The following lemma can be easily observed from the
inference rules defined above.

\begin{lem}\label{lem:superoperator} Suppose
$<P;\bar{q}=\rho>\rto{\alpha}\mu$ where $\rho\in \mathcal{D(H)}$.
Then
\begin{enumerate}
\item
if $\alpha=\qc c?r:\sigma$ for some $\qc c\in qChan$, $r\not\in
\bar{q}$, and $\sigma\in \dh$, then there exists $P'\in qProc$ such
that for any $\rho'\in \mathcal{D(H)}$,
$<P;\bar{q}=\rho'>\rto{\alpha}<P';r,\bar{q}=\sigma\otimes \rho'>$ ,

\item if $\alpha$ is not of the form $\qc c?r:\sigma$, then there
exist an index set $I$, a set of quantum processes $\{P_i : i\in
I\}$ , and a set of super-operators $\{\e_i : i\in I\}$ which only
act nontrivially on $\mathcal{L}(\h_{qv(P)})$ such that for any
$\rho'\in \mathcal{D(H)}$,
$<P;\bar{q}=\rho'>\rto{\alpha}\boxplus_{i\in I}
p_i\bullet<P_i;\bar{q}=\e_i(\rho')>$. Here $\h_{qv(P)}$ denotes the
associated Hilbert space of the quantum systems in $qv(P)$.
\end{enumerate}
\end{lem}
{\it Proof.} Obvious. \hfill $\Box$

\vspace{1em}

The transition graph of a configuration is defined as usual where
each transition $\c\rto{\alpha}\boxplus_{i=1}^n p_i\bullet \c_i$  is
depicted as
\begin{figure}[h] \centering \includegraphics {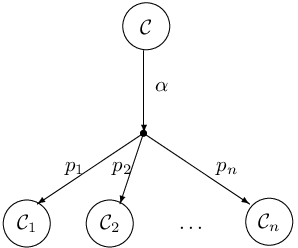}
\end{figure}\newline
and each transition of the form $\c\rto{\alpha}\d$ is simply
depicted as
\begin{figure}[h] \centering \includegraphics {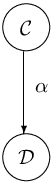}  \end{figure}

\begin{exmp}\rm
We now present a simple example to show the expressive power of our
qCCS. This example is concerned with quantum teleportation
\cite{BB93}, a famous protocol in quantum information theory which
can make use of an entangled state shared between the sender and the
receiver to teleport an unknown quantum state by sending only
classical information. This example was also considered in
\cite{JL04} and \cite{GN05}.

Let $M$ be a 2-qubit measurement such that $M=\sum_{i=0}^3
\lambda_i|\tilde{i}\>\<\tilde{i}|$, where $\tilde{i}$ is the binary
expansion of $i$. Let $CNOT$, $H$, and $\sigma_i,\ i=0,\dots,3$ be
as defined in Section 2. Then the participating quantum processes in
teleportation protocol are defined as follows:
\begin{eqnarray*}
Alice &:=& CNot[q,q_1].H[q].M[q,q_1;x].c!x.\nil, \\
Bob &:=& c?x.U_{x}[q_2].\nil,\\
Telep &:=& (Alice\| Bob)\backslash \{c\},
\end{eqnarray*}
where
\begin{eqnarray*}
U_x[q_2].\nil& := & \iif\ x=\lambda_0\ \then\ \sigma_0[q_2].\nil
\ +\ \iif\ x=\lambda_1\ \then\ \sigma_1[q_2].\nil\ +\ \\
& & \iif\ x=\lambda_2\ \then\ \sigma_3[q_2].\nil\ +\ \iif\
x=\lambda_3\ \then\ \sigma_2[q_2].\nil.
\end{eqnarray*}
The transition graph of the configuration
$$<Telep;\bar{q}=[(\alpha|0\>+\beta|1\>)\otimes\frac{1}{\sqrt{2}}(|00\>+|11\>)]>$$
is shown in Fig.\ref{fig 1 } where $\bar{q}$ is the abbreviation of
the indexed set $\{q,q_1,q_2\}$, and for any pure state $|\psi\>$,
$[|\psi\>]$ is the abbreviation of $|\psi\>\<\psi|$. Note that in
the whole procedure, Alice holds the qubits $q$ and $q_1$ while Bob
holds $q_2$. So the process $Telep$ indeed teleports the quantum
state $\alpha|0\>+\beta|1\>$ from Alice's side to Bob's side with
the aid of an EPR state.
\begin{figure}[t]\centering
\includegraphics[width=0.895\textwidth]{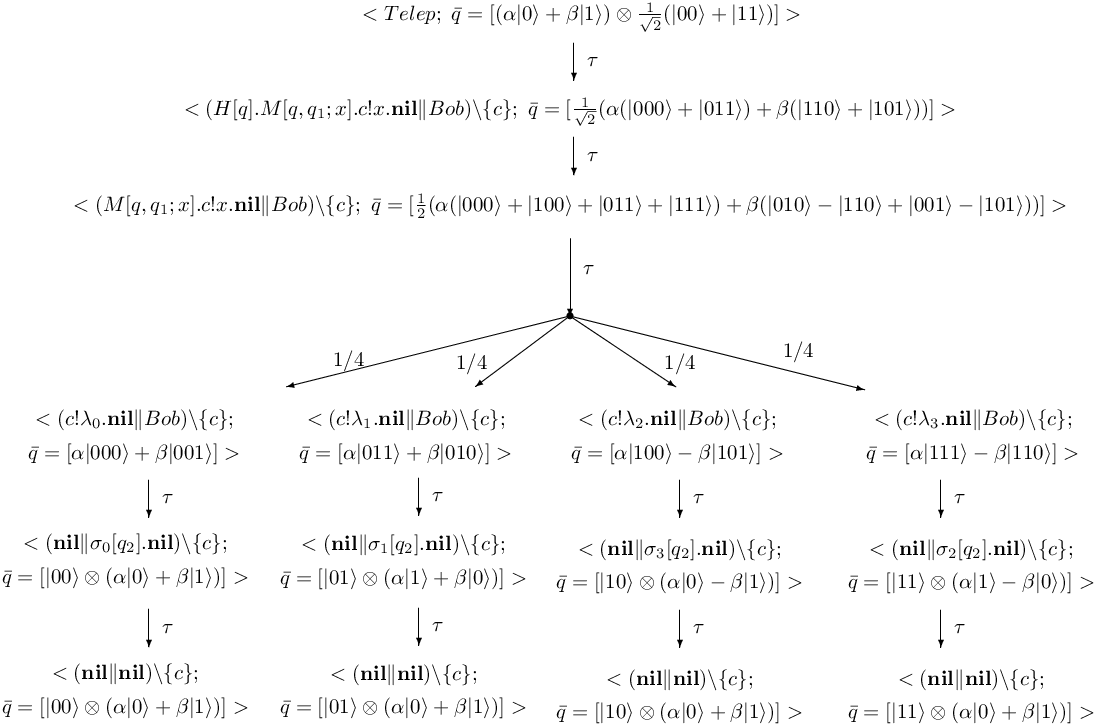}
\caption{Quantum teleportation.} \label{fig 1 }
\end{figure}
\hfill $\square$
\end{exmp}

\subsection{Combined transitions}

There are two kinds of nondeterminism in qCCS: non-probabilistic
nondeterminism caused by summation combinator `+' and probabilistic
nondeterminism caused by quantum measurements. To define
probabilistic bisimulations between quantum processes, we need a way
to resolve the first kind of nondeterminism numerically. This is
achieved in \cite{La05,La06} by treating non-probabilistic
nondeterminism as equiprobability. In this paper however, motivated
by \cite{SL94} and \cite{SL95}, we adopt a more flexible way of
allowing combining different nondeterministic choices in any
probabilistic way. To achieve this goal, a notion of adversary is
introduced. With the help of adversaries, we extend ordinary
transitions to combined transitions (resp. combined weak
transitions) which is the basis of strong probabilistic bisimulation
(resp. weak probabilistic bisimulation) defined later. Some
definitions in this subsection are motivated by or borrowed directly
from \cite{SL94} and \cite{SL95} where classical probabilistic
processes were considered.

\begin{defn}
An execution fragment $f=\c_0\alpha_1\c_1\dots\alpha_n\c_n$ is a
finite sequence of alternating configurations and actions starting
and ending with configurations, such that for each $i=0,\dots,n-1$,
there exists a transition $\c_i\rto{\alpha_{i+1}}\mu_{i+1}$ with
$\mu_{i+1}(\c_{i+1})>0$. We call $n$ the length of $f$, and denote
by $head(f)$ and $tail(f)$ the first and the last configurations of
$f$ , respectively.
\end{defn}

The set of all execution fragments is denoted by $frag$. For any
$f\in frag$, we let $Pre(f)$ be the set of execution fragments which
are prefixes of $f$.
\begin{defn}\label{def:adversary}
An adversary $\a$ is a function from execution fragments to
finite-support distributions over transitions, $i.e.$
\[
\a\ :\ frag\ \rto{}\ D(\rto{}),
\]
such that for any $f\in frag$, if $\a(f)=\boxplus_{i\in I}p_i\bullet
(\c_i, \alpha_i, \mu_i)$ then $\c_i=tail(f)$ for any $i\in I$.
\end{defn}

Intuitively, an adversary provides a mechanism to resolve
nondeterminism probabilistically by deciding next transition based
on the execution history.

\begin{defn}
Suppose $f=\c_0\alpha_1\c_1\dots\alpha_n\c_n$ is an execution
fragment and $\a$ is an adversary. We say that $f$ coincides with
$\a$ if for any $i=0,\dots,n-1$,
$\a(\c_0\alpha_1\c_1\dots\alpha_i\c_i)=\boxplus_{j\in J}p_j\bullet
(\c_i, \beta_j, \mu_j)$  such that the set $J_i=\{j\in J\ |\
\beta_{j}=\alpha_{i+1}\mbox{ and }\mu_{j}(\c_{i+1})>0\}$ is
nonempty.

We denote by $P_\a^i(f)=\sum_{j\in J_i}p_{j}\mu_{j}(\c_{i+1})$ the
probability of the $i$-th choice in $f$ according to the adversary
$\a$.
\end{defn}

For any adversary $\a$, let $F_{\c\rto{}\d}^\a$ be the set of
execution fragments with head $\c$ and tail $\d$ which coincide with
$\a$. If $f=\c_0\alpha_1\c_1\dots\alpha_n\c_n\in F_{\c\rto{}\d}^\a$,
then we denote by
\[ P_\a(f)=\prod_{i=0}^{n-1} P_\a^i(f)
\] the probability of the execution fragment $f$ according to $\a$. When
$f$ does not coincide with $\a$, we simply let $P_\a(f)=0$.

With the above definitions, we are now ready to define the notions
of combined transitions.
\begin{defn}\label{def:wtran}
For any $\c\in Con$, $s=\alpha_1\dots\alpha_n\in Act^*$, and $\mu\in
D(Con)$, we say that $\c$ can evolve into $\mu$ by a combined (resp.
a combined weak) $s$-transition, denoted by $\c\srto{s}\mu$ (resp.
$\c\Rto{s}\mu$), if there exists an adversary $\a$ such that for any
$\d\in supp(\mu)$,
\begin{enumerate}
\item
$\displaystyle\sum_{f\in F_{\c\rto{}\d}^\a} P_\a(f)=\mu(\d)$,
\item
for any $f=\c_0\beta_1\c_1\dots\beta_m\c_m\in F_{\c\rto{}\d}^\a$,
the string $\beta_1\dots\beta_m=s$ (resp. $\beta_1\dots\beta_m$ has
the form $\tau^*\alpha_1\tau^*\dots\tau^*\alpha_n\tau^*$).
\end{enumerate}
\end{defn}

In the following, we prove two lemmas which are useful for the next
sections. The first lemma shows that any convex combination of
combined $s$-transitions is also a combined $s$-transition.

\begin{lem}\label{lem:comrto}
For any $\mu_1,\dots,\mu_n\in D(Con)$ and $p_1,\dots,p_n\in (0,1)$
such that $\c\srto{s}\mu_i$ (resp. $\c\Rto{s}\mu_i$) and $\sum_i
p_i=1$, we have $\c\srto{s}\mu$ (resp. $\c\Rto{s}\mu$) for
$\mu=\sum_i p_i \mu_i$.
\end{lem}
{\it Proof.} We only prove the result for combined weak transitions
in the case of $n=2$. The general case can be proved similarly by
induction.

Suppose an adversary corresponding to $\c \Rto{s} \mu_i$ is $\a_i$,
$i=1,2$. We construct a new adversary $\a$, which will be proven to
be a corresponding adversary of $\c \Rto{s} \mu$, as follows. For
any $f\in frag$,
\begin{equation}\label{eqn:newad}
\a(f)=\left\{
\begin{array}{ll}
\displaystyle\frac{ pP_{\a_1}(f)}{ P_{\a}(f)}\a_1(f) +
(1-\frac{pP_{\a_1}(f)}{\displaystyle P_{\a}(f)})\a_2(f) \qquad & \mbox{if } P_\a(f)\neq 0, \\
p\a_1(f) + (1-p)\a_2(f) \qquad & \mbox{otherwise}.
\end{array}
\right.
\end{equation}

Note that $P_\a(\c)=1$ for any adversary $\a$ and any $\c\in Con$,
and $P_\a(f)$ is dependent only on the set $\{\a(f')\ |\ f'\in
Pre(f), \ f'\neq f\}$. The definition Eq.(\ref{eqn:newad}) is
meaningful and is an inductive one. Now we show that for any $f\in
frag$ with $head(f)=\c$,
\begin{equation}\label{eqn:paf}
P_\a(f)=pP_{\a_1}(f)+(1-p)P_{\a_2}(f)
\end{equation}
by induction on the structure of $f$.

When $f=\c$, we have
\[
P_\a(\c)=1=p+(1-p)=pP_{\a_1}(\c)+(1-p)P_{\a_2}(\c).
\]
Now suppose Eq.(\ref{eqn:paf}) holds for
$f=\c\alpha_1\c_1\dots\alpha_n\c_n$. Then for
$f'=\c\alpha_1\c_1\dots\alpha_{n+1}\c_{n+1}$, there are two cases to
consider.

{\renewcommand{\theenumi}{(\roman{enumi})}
\begin{enumerate}
\item
$P_\a(f)=0$. Then from Eq.(\ref{eqn:paf}) we also find that
$P_{\a_1}(f)=P_{\a_2}(f)=0$. So we have
$P_{\a}(f')=P_{\a_1}(f')=P_{\a_2}(f')=0$, and Eq.(\ref{eqn:paf})
holds trivially for $f'$.

\item $P_\a(f)\neq 0$. In this case, we derive that
\begin{eqnarray*}
P_\a(f')&=&P_\a(f)P_\a^{n}(f')\hspace{20em} \mbox{Definition}\\ &
=&P_\a(f) \left[\frac{pP_{\a_1}(f)}{P_{\a}(f)}P_{\a_1}^n(f') +
(1-\frac{pP_{\a_1}(f)}{P_{\a}(f)})P_{\a_2}^n(f')\right]\hspace{3.8em} \mbox{Eq.(\ref{eqn:newad})}\\
&=&pP_{\a_1}(f)P_{\a_1}^n(f') + (P_\a(f)-pP_{\a_1}(f))P_{\a_2}^n(f')\\
&=&pP_{\a_1}(f)P_{\a_1}^n(f') + (1-p)P_{\a_2}(f)P_{\a_2}^n(f')\hspace{8.3em} \mbox{Eq.(\ref{eqn:paf})}\\
&=&pP_{\a_1}(f') + (1-p)P_{\a_2}(f').\hspace{14.2em}
\mbox{Definition}
\end{eqnarray*}
\end{enumerate}
}

So for any $\d\in supp(\mu)$,
\begin{eqnarray*}
\sum_{f\in F_{\c\rto{}\d}^\a} P_\a(f)&=&\sum_{f\in
F_{\c\rto{}\d}^\a} [pP_{\a_1}(f)+(1-p)P_{\a_2}(f)]\\&=&p\sum_{f\in
F_{\c\rto{}\d}^{\a_1}} P_{\a_1}(f)+(1-p)\sum_{f\in
F_{\c\rto{}\d}^{\a_2}}P_{\a_2}(f)\\
&=&p\mu_1(\d)+(1-p)\mu_2(\d)\\
&=&\mu(\d).
\end{eqnarray*}
Here for the second equality, we have used the fact
\begin{equation}\label{eqn:adver}
F_{\c\rto{}\d}^\a=F_{\c\rto{}\d}^{\a_1}\cup F_{\c\rto{}\d}^{\a_2}
\end{equation}
which is direct from Eq.(\ref{eqn:paf}) and the observation that
$f\in F_{\c\rto{}\d}^\a$ if and only if $P_\a(f)>0$.

Furthermore, from Eq.(\ref{eqn:adver}) we deduce that for each
$f=\c_0\beta_1\c_1\dots\beta_m\c_m\in F_{\c\rto{}\d}^\a$, the string
$\beta_1\dots\beta_m$ has the form
$\tau^*\alpha_1\tau^*\dots\tau^*\alpha_n\tau^*$ since any execution
fragment in $F_{\c\rto{}\d}^{\a_1}$ and $F_{\c\rto{}\d}^{\a_2}$
does.
 \hfill
$\square$

\begin{lem}\label{lem:decRto}
Suppose $\c\srto{s}\mu$ (resp. $\c\Rto{s}\mu$),
$s=\alpha_1\dots\alpha_n\in Act^*$, and $\a$ is a corresponding
adversary. Let $\a(\c)=\boxplus_{i\in
I}p_i\bullet(\c,\beta_i,\mu_i)$. Then for any $i\in I$,
\begin{enumerate}
\item
$\beta_i=\alpha_1$ (resp. $\beta_i=\tau$ or $\alpha_1$),
\item
for any $\c'\in supp(\mu_i)$, there exist $\mu_{\c'}$ and $s'$ such
that $\c'\srto{s'}\mu_{\c'}$ (resp. $\c'\Rto{s'}\mu_{\c'}$) and
$\beta_is'=s$ (resp. $\widehat{\beta_is'}=\widehat{s}$. Here for any
$s\in Act^*$, $\widehat{s}$ denotes the string obtained from $s$ by
deleting all the occurrences of $\tau$),
\item
$\displaystyle\mu=\sum_{i\in I}\sum_{\c'\in
supp(\mu_i)}p_i\mu_i(\c')\mu_{\c'}$.
\end{enumerate}
\end{lem}
{\it Proof.} We only prove the result for combined weak transitions.
(1) is obvious. To prove (2), for any $\c'\in supp(\mu_i)$, let
$$J_{\c'}=\{j\in I\ |\ \beta_j=\beta_i\mbox{ and } \mu_j(\c')>0\},$$ $r_{\c'}=\sum_{j\in J_{\c'}}p_j \mu_j(\c')$, and
$\mu_{\c'}\in D(Con)$ such that for any $\d\in Con$,
\[
\mu_{\c'}(\d)= \frac{1}{r_{\c'}}\sum \{|\ P_\a(f)\ |\ f\in
F_{\c\rto{}\d}^\a\mbox{ and  } \c\beta_i\c'\in Pre(f) \ |\}.
\]
Here $\{|\dots|\}$ stands for the multi-set brackets. Let $s'=s$ or
$\alpha_2\dots\alpha_n$ depending on whether $\beta_i=\tau$ or
$\alpha_1$. Then $\widehat{\beta_is'}=\widehat{s}$ as required. We
now prove $\c'\Rto{s'}\mu_{\c'}$ by constructing a corresponding
adversary $\a_{\c'}$ as follows. For any $f\in frag$, let
\[
\a_{\c'}(f)=\left\{
\begin{array}{ll}
\a(\c\beta_if) \qquad  & \mbox{if } head(f)= \c', \\ 
\a(f)\qquad & \mbox{otherwise}.
\end{array}
\right.
\]
Then when $head(f)=\c'$, we have
$P_\a(\c\beta_if)=r_{\c'}P_{\a_{\c'}}(f)$. Thus for any $\d\in
supp(\mu_{\c'})$,
\begin{eqnarray*}
\sum_{f\in F_{\c'\rto{}\d}^{\a_{\c'}}} P_{\a_{\c'}}(f)&=& \frac{1}{\
r_{\c'}}\sum_{f\in F_{\c'\rto{}\d}^{\a_{\c'}}}
P_\a(\c\beta_if)\\
&=&\frac{1}{\ r_{\c'}}\sum \{|\ P_\a(f')\ |\ f'\in
F_{\c\rto{}\d}^\a\mbox{ and } \c\beta_i\c'\in Pre(f')\ |\}\\
\\
&=&\mu_{\c'}(\d).
\end{eqnarray*}

Finally, to prove (3), we need only to check that for any $\d\in
Con$,
\begin{eqnarray*}
\mu(\d)&=&\sum_{f\in F_{\c\rto{}\d}^{\a}} P_{\a}(f)\\
&=&\sum_{\c'\in \cup_i supp(\mu_i)}\sum_{i\in I_{\c'}}\sum \{|\
P_\a(f)\ |\ f\in
F_{\c\rto{}\d}^\a\mbox{ and  }\c\beta_i\c'\in Pre(f)\ |\}\\
&=&\sum_{\c'\in \cup_i supp(\mu_i)}\sum_{i\in I_{\c'}}r_{\c'} \mu_{\c'}(\d)\\
&=&\sum_{\c'\in \cup_i supp(\mu_i)}\sum_{i\in I_{\c'}} \sum_{j\in J_{\c'}}p_j \mu_j(\c') \mu_{\c'}(\d)\\
&=&\sum_{j\in I}\sum_{\c'\in supp(\mu_j)}p_j\mu_j(\c')\mu_{\c'}(\d)
\end{eqnarray*}
where $I_{\c'}=\{i\in I: \mu_i(\c')>0\}$. \hfill $\square$

\vspace{1em} To illustrate the definitions and lemmas in this
subsection, we present a simple example as follows.

\begin{exmp}\label{exam:3.2}\rm Suppose $M_{0,1}=\lambda_0|0\>\<0|+\lambda_1|1\>\<1|$
is a one-qubit measurement according to the computational basis, $H$
is the Hadarmard transformation, and
$|\pm\>=(|0\>\pm|1\>)/\sqrt{2}$. Let
\[
P= M_{0,1}[q;x].H[q].\qc c!q.\nil + \qc c!q.\nil
\]
be a quantum process which can either perform sequentially the
measurement $M$ and the transformation $H$ on $q$ before outputing
$q$, or output $q$ directly. Now consider the configuration
\[ \c=<P;q=|+\>\<+|>.\] The transition graph of $\c$ can be depicted as in Fig.~\ref{fig:tmp}
\begin{figure}[t] \centering \includegraphics{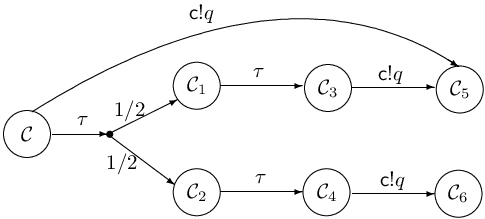}
\caption{Transition graph of $\c$ in Example~\ref{exam:3.2}.\label{fig:tmp}}
\end{figure}
where
\[\begin{array}{ll}
\c_1 = <H[q].\qc c!q.\nil; q=|0\>\<0|>,\ \  &\c_2 = <H[q].\qc
c!q.\nil; q=|1\>\<1|>,\\ \c_3 = <\qc c!q.\nil; q=|+\>\<+|>,\ \
&\c_4 = <\qc c!q.\nil; q=|-\>\<-|>,\\\c_5 = <\nil; q=|+\>\<+|>,\
\ &\c_6 = <\nil; q=|-\>\<-|>.
\end{array}
\]
Then by taking an adversary $\a_1$ such that
\[
\begin{array}{ll}
 \a_1(\c)=(\c,\tau,\displaystyle\frac{1}{2}\bullet \c_1 \boxplus
\frac{1}{2}\bullet \c_2),\ \ \ \ &\a_1(\c\tau\c_1)=
(\c_1,\tau,\c_3), \\\a_1(\c\tau\c_2)=(\c_2,\tau, \c_4),\ \ \ \
&\a_1(\c\tau\c_1\tau\c_3)= (\c_3,\qc c!q,\c_5),
\end{array}
\] and $$\a_1(\c\tau\c_2\tau\c_4)= (\c_4,\qc c!q,\c_6),$$
 we have the combined (weak) transitions
\[\c\srto{\tau\tau\qc c!q} \frac{1}{2}\bullet\c_5\boxplus
\frac{1}{2}\bullet \c_6 \hspace{2em}\mbox{ and }
\hspace{2em}\c\Rto{\qc c!q} \frac{1}{2}\bullet\c_5\boxplus
\frac{1}{2}\bullet \c_6.\] On the other hand, the adversary $\a_2$
satisfying $\a_2(\c)= (\c,\qc c!q,\c_5)$ leads to the combined weak
transition $\c\Rto{\qc c!q} \c_5$. Thus for any $p\in [0,1]$, we
have
\[\c\Rto{\qc c!q} (1-\frac{p}{2})\bullet\c_5\boxplus
\frac{p}{2}\bullet \c_6\] by combining the above two weak $\qc
c!q$-transitions. The corresponding adversary $\a$ is constructed as
\begin{eqnarray*}
\a(\c)&=&p \a_1(\c) + (1-p) \a_2(\c) =
p\bullet(\c,\tau,\frac{1}{2}\bullet \c_1 \boxplus
\frac{1}{2}\bullet \c_2)\boxplus (1-p)\bullet(\c,\qc c!q,\c_5),
\\\a(\c\tau\c_1)&=& \a_1(\c\tau\c_1)=(\c_1,\tau,\c_3),\ \ \ \cdots.
\end{eqnarray*}
 \hfill $\square$
\end{exmp}

\section{Strong probabilistic bisimulation between quantum processes}

This section is devoted to the notion of strong probabilistic
bisimulation between quantum processes and its properties such as
congruence under various combinators.

Given an equivalence relation $\r\subseteq Con\times Con$, two
distributions $\mu$ and $\nu$ on $Con$ are said to be equivalent
under $\r$, denoted by $\mu\equiv_\r \nu$, if for any equivalence
class $M\in Con/\r$ it holds $\mu(M)=\nu(M)$. Two quantum contexts
$\bar{q}=\rho$ and $\bar{r}=\sigma$ are equal if there exists a
permutation $\Pi$ such that $\Pi(\bar{q})=\bar{r}$ and at the same
time $\Pi\rho\Pi^\dag=\sigma$. We denote $\c\nrto{\alpha}$ if there
exists no $\mu\in D(Con)$ such that $\c\rto{\alpha}\mu$; we simply
write $\c\nrto{}$ if $\c\nrto{\alpha}$ for all $\alpha\in Act$.

\begin{defn}\label{def:bisimulation}
An equivalence relation $\r\subseteq Con\times Con$ is a strong
probabilistic bisimulation if for any $\c,\d\in Con$, $(\c,\d)\in
\r$ implies that
\begin{enumerate}
\item whenever $\c \rto{\alpha} \mu$  for some $\alpha$ and $\mu$, there
exists $\nu$ such that $\d \srto{\alpha} \nu$ and $\mu\equiv_\r\nu$,
\item
if $\c\nrto{}$, then $Contex(\c)=Contex(\d)$.
\end{enumerate}
\end{defn}

As mentioned in Section 1, one of the purposes of qCCS is to provide
a theoretical framework to describe quantum concurrent systems such
as quantum cryptographic protocols. As a consequence, not only the
observable actions but also the quantum operations such as unitary
transformations and measurements performed by processes must be
taken into consideration when bisimulation relations are
investigated. For example, we cannot in any sense regard a quantum
process which can merely sequentially perform 5 $\tau$ actions and
then terminates as bisimilar to the teleportation process $Telep$
defined in Example 3.1. Furthermore, because of the possible
entanglement between different quantum systems, the effect of
quantum operations can be fully reflected only by state change of
the whole quantum context. This is the reason why we need clause (2)
in Definition \ref{def:bisimulation}. The clause (1) is originated
from \cite{LS91} and \cite{SL94}.

\begin{defn}\begin{enumerate} \item
Two configurations $\c$ and $\d$ are strongly bisimilar, denoted by
$\c\sim_c\d$, if there is a strong probabilistic bisimulation $\r$
such that $(\c,\d)\in \r$.
\item
Two processes $P$ and $Q$ are strongly bisimilar, denoted by
$P\sim_p Q$, if for any context $C$ and any indexed set $\bar{v}$ of
values, $<P[\bar{v}/\bar{x}];C>\sim_c < Q[\bar{v}/\bar{x}];C>$. Here
$\bar{x}$ is the set of free classical variables contained in
processes $P$ and $Q$.
\end{enumerate}
\vspace{1em} We usually omit the subscripts of $\sim_c$ and $\sim_p$
when no confusion arises.
\end{defn}

The difference between our notion of probabilistic bisimulation and
the probabilistic branching bisimulation defined in \cite{La05,La06}
can be best illustrated by the following example.
\begin{exmp}\rm Suppose $M_{0,1}$, $H$, and $|+\>$ are given
as in Example \ref{exam:3.2}, and $U=\sigma_1 H$. Suppose
\[ \c=<H[q].\nil + U[q].\nil + M_{0,1}[q;x].\nil; q=|+\>\<+|> \] and
\[
\d=<H[q].\nil + U[q].\nil; q=|+\>\<+|>\] with transition graphs
depicted as
\begin{figure}[h] \centering \includegraphics[width=.7\textwidth]{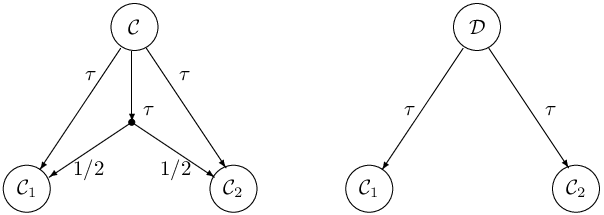}
\label{fig 4.1 }
\end{figure}
 \newline where
$$\c_1=<\nil;q=|0\>\<0|>\ \ \mbox{and}\ \ \c_2=<\nil;q=|1\>\<1|>.$$
Then $\c$ and $\d$ are bisimilar in our notion of strong
probabilistic bisimulation, since $\d$ can simulate the action
$M_{0,1}[q;x]$ of $\c$ by choosing its actions $H[q]$ and $U[q]$
with respective probabilities one half.

Note that in the sense of probabilistic branching bisimulation
presented in \cite{La05,La06}, the configurations $\c$ and $\d$ are
also bisimilar. But the reason is that state change of contexts
caused by quantum operations is not considered there. As a
consequence, the configurations $\c_1$ and $\c_2$, which are not
bisimilar in our sense of bisimulation, are treated to be bisimilar
in \cite{La05,La06}.\hfill $\square$
\end{exmp}

 In the following, we derive some properties of strong
probabilistic bisimulation. The proofs are similar to but much
simpler than those of the corresponding results for weak
probabilistic bisimulation in the next section except for Theorem
\ref{lem:sprepreserve} (2), so we omit them here.

\begin{thm}
$\sim$ is the largest strong probabilistic bisimulation on $Con$.
\end{thm}

\begin{thm}\label{lem:sbisimulation}
For any $\c,\d\in Con$, $\c\sim \d$ if and only if for any $s\in
Act^*$,
\begin{enumerate}
\item
whenever $\c \srto{s} \mu$ for some $\mu$, then there exists $\nu$
such that $\d \srto{{s}} \nu$ and $\mu\equiv_\sim\nu$,

\item whenever $\d \srto{s} \nu$ for some $\nu$, then there exists $\mu$ such that $\c
\srto{s} \mu$ and $\mu\equiv_\sim\nu$,

\item if $\c\nrto{}$ and $\d\nrto{}$, then $Contex(\c)=Contex(\d)$.
\end{enumerate}
\end{thm}

    \begin{thm}\label{lem:sprepreserve} If $P\sim Q$ then
\begin{enumerate}
\item
$a.P\sim a.Q$, for any $a\in \{c?x,c!e,\qc c?q,\qc
c!q,U[\bar{q}],M[\bar{q};x]\}$;
\item
 $P+R\sim Q+R$ for any $R$;
\item
$P\| R\sim Q\|R$ provided that $R$ is free of unitary transformation
and measurement, or $P$ and $Q$ are free of quantum input;
\item
$P[f]\sim Q[f]$, for any relabeling function $f$;
\item
$\iif\ b\ \then\ P \sim \iif\ b\ \then\ Q$, for any boolean
expression $b$.
\end{enumerate}
\end{thm}
{\it Proof.} The cases other than (2) are simpler than the
counterparts for weak probabilistic bisimulation. In the following,
we prove (2) by showing a stronger result: for any contexts $C$ and
$D$, if $<P_i[\bar{v}/\bar{x}];C>\sim<Q_i[\bar{v}/\bar{x}];D>$ for
$i=1,2$, then
$<P_1[\bar{v}/\bar{x}]+P_2[\bar{v}/\bar{x}];C>\sim<Q_1[\bar{v}/\bar{x}]+Q_2[\bar{v}/\bar{x}];D>$.
Here $\bar{x}$ is the set of free classical variables contained in
processes $P_i$ and $Q_i$.

Suppose
$<P_1[\bar{v}/\bar{x}]+P_2[\bar{v}/\bar{x}];C>\rto{\alpha}\mu$ for
some $\alpha$ and $\mu$. Then from \textbf{Sum} rule, we have
$<P_1[\bar{v}/\bar{x}];C>\rto{\alpha}\mu$ or
$<P_2[\bar{v}/\bar{x}];C>\rto{\alpha}\mu$. By the assumption
$<P_i[\bar{v}/\bar{x}];C>\sim<Q_i[\bar{v}/\bar{x}];D>$ and Theorem
\ref{lem:sbisimulation}, it holds that
$<Q_1[\bar{v}/\bar{x}];D>\srto{\alpha}\nu$ or
$<Q_2[\bar{v}/\bar{x}];D>\srto{\alpha}\nu$ for some $\nu$ such that
$\mu\equiv_\sim\nu$. In either case, using \textbf{Sum} rule again,
we have
$<Q_1[\bar{v}/\bar{x}]+Q_2[\bar{v}/\bar{x}];D>\srto{\alpha}\nu$.

Similarly, if
$<Q_1[\bar{v}/\bar{x}]+Q_2[\bar{v}/\bar{x}];D>\rto{\alpha}\nu$ for
some $\alpha$ and $\nu$, we can also find a $\mu$ such that
$<P_1[\bar{v}/\bar{x}]+P_2[\bar{v}/\bar{x}];C>\srto{\alpha}\mu$ and
$\mu\equiv_\sim \nu$.

Finally, if $<P_1[\bar{v}/\bar{x}]+P_2[\bar{v}/\bar{x}];C>\nrto{}$
and $<Q_1[\bar{v}/\bar{x}]+Q_2[\bar{v}/\bar{x}];D>\nrto{}$, then we
have $<P_1[\bar{v}/\bar{x}];C>\nrto{}$ and
$<Q_1[\bar{v}/\bar{x};D]>\nrto{}$. Hence $C=D$ from the assumption
that $<P_1[\bar{v}/\bar{x}];C>\sim<Q_1[\bar{v}/\bar{x}];D>$. Then
the result follows from Theorem \ref{lem:sbisimulation}.
 \hfill $\Box$
\begin{thm} For any $P,Q,R\in qProc$,
\begin{enumerate}
\item
$P+\nil\sim P$,
\item
$P+P\sim P$,
\item
$P+Q\sim Q+P$,
\item
$P+(Q+R)\sim(P+Q)+R$,
\item
$P\|\nil \sim P$,
\item
$P\|Q \sim Q\|P$,
\item
$P\| (Q\| R)\sim (P\| Q)\| R$.
\end{enumerate}
\end{thm}

\section{Weak probabilistic bisimulation between quantum processes}

As in classical CCS, the notion of weak probabilistic bisimulation
which abstracts from unobservable internal actions is more useful in
implementation and verification. In this section, based on the
notion of combined weak transition introduced in Section 3.3, we
present weak probabilistic bisimulation for our qCCS.

\begin{defn}\label{def:wbisimulation}
An equivalence relation $\r\subseteq Con\times Con$ is a weak
probabilistic bisimulation if for any $\c,\d\in Con$, $(\c,\d)\in
\r$ implies that
\begin{enumerate}
\item
whenever $\c \rto{\alpha} \mu$ for some $\alpha$ and $\mu$, there
exists $\nu$ such that $\d \Rto{\widehat{\alpha}} \nu$ and
$\mu\equiv_\r\nu$,
\item
if $\c\nrto{}$ and $\d\nrto{}$, then $Contex(\c)=Contex(\d)$.
\end{enumerate}
\end{defn}

The following lemma shows that the ordinary transition in clause (1)
of the above definition can be strengthened to combined weak
transition.

\begin{lem}\label{lem:Rtotran} Let $\r\subseteq Con\times Con$ be a weak
probabilistic bisimulation and $(\c,\d)\in \r$. Then for any $s\in
Act^*$, if $\c \Rto{s} \mu$, then $\d \Rto{\widehat{s}} \nu$ for
some $\nu$ such that $\mu\equiv_\r\nu$.
\end{lem}
{\it Proof.} Let $\a$ be an adversary corresponding to
$\c\Rto{s}\mu$. Since there are no recursive constructs in qCCS, we
can prove this lemma by induction on the maximal length $h$ of the
execution fragments in $\cup_{\d\in supp(\mu)}F_{\c\rto{}\d}^\a$.

If $h=0$, then $s$ is the empty string and $\mu=\c$. In this case,
we need only to take $\nu=\d$.

Suppose the result holds for $h\leq n$. We now prove that it also
holds for $h=n+1$. Let $\a(\c)=\boxplus_{i\in
I}p_i\bullet(\c,\alpha_i,\mu_i)$. Then for each $i\in I$ we have
$\c\rto{\alpha_i}\mu_i$, and so there exists $\nu_i$ such that
$\d\Rto{\widehat{\alpha_i}}\nu_i$ and $\mu_i\equiv_\r \nu_i$.
Furthermore, from Lemma \ref{lem:decRto}, for any $\c'\in
supp(\mu_i)$ there exist $\mu_{\c'}$ and $s'$ such that
$\c'\Rto{s'}\mu_{\c'}$, $\widehat{\alpha_is'}=\widehat{s}$, and
\[
\mu=\sum_{i\in I}\sum_{\c'\in supp(\mu_i)}p_i\mu_i(\c')\mu_{\c'}.
\]
Now take arbitrarily $\d'\in supp(\nu_i)$. Let $[\d']_\r$ denote the
equivalence class of $\r$ in which $\d'$ lies. Then $supp(\mu_i)\cap
[\d']_\r\neq \emptyset$ from $\mu_i\equiv_\r \nu_i$. For any $\c'\in
supp(\mu_i)\cap [\d']_\r$, we can choose an adversary $\a_{\c'}$
corresponding to $\c'\Rto{s'}\mu_{\c'}$ such that the maximal length
of the execution fragments in $\cup_{\d\in
supp(\mu_{\c'})}F_{\c'\rto{}\d}^{\a_{\c'}}$ is less than $n+1$. So
by induction we have $\d'\Rto{\widehat{s'}}\nu_{\d'}^{\c'}$ for some
$\nu_{\d'}^{\c'}$, and $\mu_{\c'}\equiv_\r\nu_{\d'}^{\c'}$. From
Lemma \ref{lem:comrto} it holds $\d'\Rto{\widehat{s'}} \nu_{\d'}$
where
\[
\nu_{\d'}= \sum_{\c'\in supp(\mu_i)\cap [\d']_\r}
\frac{\mu_i(\c')}{\mu_i(supp(\mu_i)\cap [\d']_\r)}\nu_{\d'}^{\c'}.
\]
It is now direct to check that $\d\Rto{\widehat{s}}\nu$ for
\[
\nu= \sum_{i\in I}\sum_{\d'\in supp(\nu_i)}p_i\nu_i(\d')\nu_{\d'}.
\]

Finally, we show that $\mu\equiv_\r\nu$. For any $M\in Con/\r$,
\begin{eqnarray*}
\nu(M)&=&\sum_{i\in I}\sum_{\d'\in
supp(\nu_i)}p_i\nu_i(\d')\nu_{\d'}(M)\\
&=&\sum_{i\in I}\sum_{\d'\in supp(\nu_i)}p_i\nu_i(\d')\sum_{\c'\in
supp (\mu_i)\cap [\d']_\r}
\frac{\mu_i(\c')}{\mu_i(supp(\mu_i)\cap [\d']_\r)}\nu_{\d'}^{\c'}(M)\\
&=&\sum_{i\in I}\sum_{\c'\in
supp(\mu_i)}p_i\mu_i(\c')\mu_{\c'}(M)\sum_{\d'\in supp(\nu_i)\cap
[\c']_\r}
\frac{\nu_i(\d')}{\mu_i(supp(\mu_i)\cap [\c']_\r)}\\
&=&\sum_{i\in I}\sum_{\c'\in supp(\mu_i)}p_i\mu_i(\c')\mu_{\c'}(M)
\frac{\nu_i([\c']_\r)}{\mu_i([\c']_\r)}\\
&=&\sum_{i\in I}\sum_{\c'\in
supp(\mu_i)}p_i\mu_i(\c')\mu_{\c'}(M)\\
&=&\mu(M).
 \end{eqnarray*}
Here the third equality is due to the fact that
$\mu_{\c'}\equiv_\r\nu_{\d'}^{\c'}$ for any $\d'\in supp(\nu_i)$ and
$\c'\in supp(\mu_i)\cap [\d']_\r$; the fifth equality holds because
$\mu_i\equiv_\r \nu_i$ for any $i\in I$.  \hfill $\square$

\begin{lem}\label{lem:Rto}
Let $\r\subseteq Con\times Con$ be a weak probabilistic bisimulation
and $(\c,\d)\in \r$.
\begin{enumerate}
\item
If $\c\nrto{}$ then $\d\nrto{\alpha}$ for any $\alpha\in
Act-\{\tau\}$.
\item
For any $s\in Act^*$, if $\c\Rto{s}\mu$ such that $\c'\nrto{}$ for
some $\c'\in supp(\mu)$, then there exists $\nu$ such that
$\d\Rto{\widehat{s}}\nu$ and $\d'\nrto{}$ for some $\d'\in
supp(\nu)$. Furthermore, $Contex(\c')=Context(\d')$.
\end{enumerate}
\end{lem}
{\it Proof.} (1) is easy. To prove (2), from $\c\Rto{s}\mu$ we first
find some $\nu_1$ such that $\d\Rto{\widehat{s}}\nu_1$ and
$\mu\equiv_\r\nu_1$. If there exists a $\d_1\in
supp(\nu_1)\cap[\c']_\r$ such that $\d_1\nrto{}$ then we are done.
Otherwise, for any $\d_1\in supp(\nu_1)\cap[\c']_\r$, from
$\c'\nrto{}$ and (1) we have $\d_1\rto{{\tau}}\nu_2$ for some
$\nu_2$ such that $\c'\r\d_2$ for any $\d_2\in supp(\nu_2)$. Then we
check if there exists a $\d_2\in supp(\nu_2)$ such that
$\d_2\nrto{}$. Note that the quantum processes we consider in this
paper are all finitely derivable. It follows that we will finally
find a distribution $\nu$ such that $\d\Rto{\widehat{s}}\nu$ and
there exists some $\d'\in supp(\nu)$ satisfying $\c'\r\d'$ and
$\d'\nrto{}$. Furthermore, from Definition \ref{def:wbisimulation}
(2) we have $Context(\c')=Context(\d')$.
 \hfill
$\square$

\vspace{1em} Since the union of equivalence relations is not
necessarily an equivalence relation, the union of weak probabilistic
bisimulations is not necessarily a weak probabilistic bisimulation
either. Nevertheless, we can prove that the reflexive and transitive
closure of the union of weak probabilistic bisimulations is also a
weak probabilistic bisimulation.

\begin{thm}\label{thm:lbisimulation}
If $\r_i, i\in I$, is a collection of weak probabilistic
bisimulations on $Con$, then their reflexive and transitive closure
$(\cup_i \r_i)^*$ is also a weak probabilistic bisimulation.
\end{thm}
{\it Proof.} By definition, $\r_i$ is symmetric for any $i\in I$. So
$(\cup_i \r_i)^*$ is also symmetric and hence an equivalence
relation. Now suppose $(\c,\d)\in (\cup_i \r_i)^*$. Then there exist
an integer $n$ and a series of configurations $\c_0,\dots,\c_n$ such
that $\c_0=\c$, $\c_n=\d$, and $(\c_i,\c_{i+1})\in \r_{k_i}$ for
some $k_i\in I$, $i=0,\dots,n-1$. There are two cases we should
consider: {\renewcommand{\theenumi}{(\roman{enumi})}
\begin{enumerate}
\item
$\c\rto{\alpha}\mu_0$ for some $\alpha$ and $\mu_0$. Then from $\c
\r_{k_0} \c_1$, there exists $\mu_1$ such that
$\c_1\Rto{\widehat{\alpha}}\mu_1$ and $\mu_0(M_0)=\mu_1(M_0)$ for
any $M_0\in Con/\r_{k_0}$. Furthermore, from $\c_1 \r_{k_1} \c_2$
and Lemma \ref{lem:Rtotran}, we have
$\c_2\Rto{\widehat{\alpha}}\mu_2$ for some $\mu_2$, and
$\mu_1(M_1)=\mu_2(M_1)$ for any $M_1\in Con/\r_{k_1}$. In this way,
we can derive that $\c_{i+1}\Rto{\widehat{\alpha}}\mu_{i+1}$ for
some $\mu_{i+1}$ such that $\mu_i(M_i)=\mu_{i+1}(M_i)$ for any
$M_i\in Con/\r_{k_i}$, $i=0,\dots,n-1$. Now suppose $M\in
Con/(\cup_i \r_i)^*$. Notice that for any $i=0,\dots,n-1$, $M$ is
the disjoint union of some equivalence classes of $Con/\r_{k_i}$
since $\r_{k_i}\subseteq (\cup_i \r_i)^*$. It follows that
$\mu_i(M)=\mu_{i+1}(M)$ for any $i=0,\dots,n-1$. Thus we have
$\mu_0(M)=\mu_n(M)$.

\item $\c\nrto{}$ and $\d\nrto{}$. Then from $\c \r_{k_0} \c_1$ and Lemma
\ref{lem:Rto} we have $\c_1\Rto{\widehat{\tau}}\mu_1$, and there
exists some $\d_1\in supp(\mu_1)$ such that $\d_1\nrto{}$ and
$Contex(\c)=Context(\d_1)$. Similarly, for any $i=2,\dots,n$ we can
derive that $\c_i\Rto{\widehat{\tau}}\mu_i$, and there exists some
$\d_i\in supp(\mu_i)$ such that $\d_i\nrto{}$ and
$Contex(\d_{i-1})=Context(\d_i)$. Finally, from the fact
$\d\nrto{}$, it is the only case that $\d_n=\d$ and so
$Context(\d)=Context(\d_{n-1})=\dots=Context(\c)$.
\end{enumerate}
} From (i) and (ii), we know that $(\cup_i \r_i)^*$ is also a weak
probabilistic bisimulation. \hfill $\square$

\begin{defn}\begin{enumerate} \item
Two configurations $\c$ and $\d$ are weakly bisimilar, denoted by
$\c\approx_c\d$, if there is a weak probabilistic bisimulation $\r$
such that $(\c,\d)\in \r$.

\item Two quantum processes $P$ and $Q$ are weakly bisimilar,
denoted by $P\approx_p Q$, if for any context $C$ and any indexed
set $\bar{v}$ of values, $<P[\bar{v}/\bar{x}];C>\approx_c <
Q[\bar{v}/\bar{x}];C>$. Here $\bar{x}$ is the set of free classical
variables contained in processes $P$ and $Q$.
\end{enumerate}
\vspace{1em} We usually omit the subscripts of $\approx_c$ and
$\approx_p$ when no confusion arises.
\end{defn}

We now show that the weak bisimilarity relation $\approx$ is a weak
probabilistic bisimulation; it is in fact the largest weak
probabilistic bisimulation on $Con$.

\begin{cor}\label{cor:wbisimilation}
$\approx$ is a weak probabilistic bisimulation on $Con$.
\end{cor}
{\it Proof.} By definition, we have
$$\approx=\bigcup\{\r\ | \ \r \mbox{ is a weak probabilistic bisimulation on }Con\}.$$ From Theorem \ref{thm:lbisimulation},
the reflexive and transitive closure $\approx^*$ is also a weak
probabilistic bisimulation. Hence $\approx^*\subseteq \approx$. On
the other hand, we have obviously $\approx\subseteq \approx^*$. So
we derive that $\approx= \approx^*$, and then $\approx$ is also a
weak probabilistic bisimulation. \hfill $\square$

\vspace{1em}

The next theorem gives us a necessary and sufficient condition to
decide whether a pair of configurations are weakly bisimilar.

\begin{thm}\label{thm:bisimulation}
For any $\c,\d\in Con$, $\c\approx \d$ if and only if for any $s\in
Act^*$,
\begin{enumerate}
\item
whenever $\c \Rto{s} \mu$ then there exists $\nu$ such that $\d
\Rto{\widehat{s}} \nu$ and $\mu\equiv_\approx\nu$,
\item
whenever $\d \Rto{s} \nu$ then there exists $\mu$ such that $\c
\Rto{\widehat{s}} \mu$ and $\mu\equiv_\approx\nu$,
\item
if $\c\nrto{}$ and $\d\nrto{}$, then $Contex(\c)=Contex(\d)$.
\end{enumerate}
\end{thm}
{\it Proof.} First, we define a new relation $\approx'$ on $Con$
such that $\c\approx'\d$ if and only if for any $s\in Act^*$, the
conditions (1), (2), and (3) hold. It is obvious that $\approx'$ is
an equivalence relation. Furthermore, from Corollary
\ref{cor:wbisimilation} and Lemma \ref{lem:Rtotran}, we have
$\approx\subseteq \approx'$. Then $\approx'$ is also a weak
probabilistic bisimulation on $Con$ since $\mu\equiv_\approx\nu$
implies $\mu\equiv_{\approx'}\nu$. Hence  we have $\approx'\subseteq
\approx$ and then $\approx = \approx'$. \hfill $\square$

\subsection{Congruence of weak probabilistic bisimilarity}
This subsection is devoted to the congruence property of weak
probabilistic bisimilarity.
\begin{lem}\label{lem:t}
If $P\approx Q$, then $P[r/q]\approx Q[r/q]$ for any $r\not \in
qv(P)\cup qv(Q)$.
\end{lem}
{\it Proof.} It is direct to check that for any quantum contexts $C$
and $D$, $<P[r/q]; C>\approx <Q[r/q]; D>$ if and only if
$<P;C[q'/q][q/r]>\approx <Q;D[q'/q][q/r]>$ where $q'\not\in
qv(C)\cup qv(D)$. Then the lemma follows.\hfill $\Box$

\begin{thm}\label{lem:wprepreserve} If $P\approx Q$ then $a.P\approx a.Q$ for any $a\in \{c?x,c!e,\qc
c?q,\qc c!q,U[\bar{r}],M[\bar{r};x]\}$.
\end{thm}
{\it Proof.} Assume that $\bar{x}$ is the set of free classical
variables contained in processes $P$ and $Q$. For any context $C$
and any indexed value set $\bar{v}$, we need to prove
$<a.P[\bar{v}/\bar{x}];C>\approx <a.Q[\bar{v}/\bar{x}];C>$. Suppose
$<a.P[\bar{v}/\bar{x}];C>\rto{\alpha}\mu$ and $C$ is of the form
$\bar{q}=\rho$. We only consider the cases where $a$ has the form
$\qc c?q$ or $M[\bar{r};x]$; other cases are simpler.
{\renewcommand{\theenumi}{(\roman{enumi})}
\begin{enumerate}
\item $a= \qc c?q$. There are two subcases to consider.

\begin{enumerate} \item
$\alpha=\qc c?r$ for some $r\in \bar{q}-qv(\qc c?q.P)$. Then
$\mu=<P[\bar{v}/\bar{x}][r/q];C>$. From \textbf{Q-Inp2} rule, we
have
$<a.Q[\bar{v}/\bar{x}];C>\rto{\alpha}<Q[\bar{v}/\bar{x}][r/q];C>$,
and furthermore, $<P[\bar{v}/\bar{x}][r/q];C>\approx
<Q[\bar{v}/\bar{x}][r/q];C>$ from the assumption that $P\approx Q$
and Lemma \ref{lem:t}.

\item $\alpha=\qc c?r:\sigma$ for some $r\not\in \bar{q}$ and $\sigma\in
\dh$. Then $\mu=<P[\bar{v}/\bar{x}][r/q];r,\bar{q}=\sigma\otimes
\rho>$. From \textbf{Q-Inp1} rule, we have
$<a.Q[\bar{v}/\bar{x}];C>\rto{\alpha}<Q[\bar{v}/\bar{x}][r/q];r,\bar{q}=\sigma\otimes
\rho>$. Furthermore, we can check that
$<P[\bar{v}/\bar{x}][r/q];r,\bar{q}=\sigma\otimes \rho>\approx
<Q[\bar{v}/\bar{x}][r/q];r,\bar{q}=\sigma\otimes \rho>$ from the
assumption that $P\approx Q$ and Lemma \ref{lem:t}.
\end{enumerate}

\item $a=M[\bar{r};x]$, $M$ has the spectral decomposition $M=\sum_i
\lambda_i P_i$. Then $\alpha=\tau$ and
$\mu=\boxplus{p_i}\bullet<P[\bar{v}/\bar{x},\lambda_i/x];\bar{q}=P_{i,\bar{r}}\rho
P_{i,\bar{r}}/p_i>$, where $p_i=\tr P_{i,\bar{r}}\rho$. From
\textbf{Meas} rule, we derive
\[
<a.Q[\bar{v}/\bar{x}];C>\rto{\alpha}
\nu=\boxplus{p_i}\bullet<Q[\bar{v}/\bar{x},\lambda_i/x];\bar{q}=P_{i,\bar{r}}\rho
P_{i,\bar{r}}/p_i>.
\]
Furthermore, for any $N\in Con/\approx$,
$$\mu(N)=\sum_i \{|\ p_i\ |\ <P[\bar{v}/\bar{x},\lambda_i/x];\bar{q}=P_{i,\bar{r}}\rho
P_{i,\bar{r}}/p_i>\in N|\}$$ and
$$\nu(N)=\sum_i \{|\ p_i\ |\ <Q[\bar{v}/\bar{x},\lambda_i/x];\bar{q}=P_{i,\bar{r}}\rho
P_{i,\bar{r}}/p_i>\in N|\}.$$ By the assumption $P\approx Q$, we
have for any context $D$, $<P[\bar{v}/\bar{x},\lambda_i/x];D>\in N$
if and only if $<Q[\bar{v}/\bar{x},\lambda_i/x];D>\in N$. Thus
$\mu(N)=\nu(N)$.
\end{enumerate}
} Symmetrically, we can prove that if
$<a.Q[\bar{v}/\bar{x}];C>\rto{\alpha}\nu$ for some $\alpha$ and
$\nu$, then there exists a transition
$<a.P[\bar{v}/\bar{x}];C>\rto{\alpha}\mu$ such that
$\mu\equiv_\approx \nu$. Then the result of this theorem holds by
using Theorem \ref{thm:bisimulation}. \hfill $\square$

\vspace{1em}

For the sake of simplicity, in the rest of this subsection we only
consider closed quantum processes. The same results can be extended
easily to the case of quantum processes with free classical
variables.

\begin{thm}\label{lem:wrelpreserve} If $P\approx Q$ then $P[f]\approx Q[f]$ for any relabeling function $f$.
\end{thm}
{\it Proof.} Let
\begin{eqnarray} \r'&=&\{(<P[f];C>, <Q[f];D>)\ \ |\
<P;C>\approx <Q;D>,\nonumber
\\
&& \hspace{17em}\mbox{ and $f$ is a relabeling function}\}
\end{eqnarray}
and $\r=(\r'\cup \approx)^*$ be the equivalence closure ($i.e.$ the
reflexive, symmetric and transitive closure) of $\r'\cup\approx$. We
prove in the following that $\r$ is a weak probabilistic
bisimulation on $Con$.

Suppose $(\c,\d)\in \r$. We may assume that $(\c,\d)\in \r'$ because
the extension to the equivalence closure is straightforward. So we
can suppose further that $\c=<P[f];C>$ and $\d= <Q[f];D>$ for some
$<P;C>\approx <Q;D>$, and $f$ is a relabeling function.

{\renewcommand{\theenumi}{(\roman{enumi})}
\begin{enumerate}
\item
If $<P[f];C>\rto{\alpha}\mu$, then by\textbf{ Rel} rule, there
exists a transition $<P;C>\rto{\beta}\mu_1=\boxplus p_i \bullet
<P_i;C_i>$ such that $\alpha=\beta[f]$ and $\mu=\boxplus p_i \bullet
<P_i[f];C_i>$. By the assumption that $<P;C>\approx <Q;D>$, we have
$<Q;D>\Rto{\widehat{\beta}}\nu_1=\boxplus q_j \bullet <Q_j;D_j>$
such that $\mu_1\equiv_\approx \nu_1$. Then by\textbf{ Rel} rule, it
holds that $$<Q[f];D>\Rto{\widehat{\alpha}}\nu=\boxplus q_j \bullet
<Q_j[f];D_j>$$ and furthermore, $\mu\equiv_\r \nu$ by the fact that
$\mu_1\equiv_\approx \nu_1$ and the definition of $\r$.

\item  If $<P[f];C>\nrto{}$ and $<Q[f];D>\nrto{}$,
then we have  $<P;C>\nrto{}$ and $<Q;D>\nrto{}$. Hence $C=D$ from
the assumption that $<P;C>\approx<Q;D>$.
\end{enumerate}
} From (i) and (ii) we know that $\r$ is a weak probabilistic
bisimulation on $Con$. Since $P\approx Q$, we have $<P;C>\approx
<Q;C>$  for any quantum context $C$, and so $(<P[f];C>, <Q[f];C>)\in
\r$. Hence $<P[f];C>\approx <Q[f];C>$, and $P[f]\approx Q[f]$ from
the arbitrariness of $C$. \hfill $\square$

\begin{thm}\label{lem:wchopreserve} If $P\approx Q$ then $\iif\ b\ \then\ P\approx \iif\ b\ \then\ Q$
 for any boolean expression $b$.
\end{thm}
{\it Proof.} Obvious.\hfill $\square$

\vspace{1em}

Theorems \ref{lem:wprepreserve} -- \ref{lem:wchopreserve} imply that
weak probabilistic bisimilarity is preserved by prefix, relabeling,
and conditional choice. However, it is not preserved by restriction.
An example is as follows. Let $U_1,U_2,V_1,V_2$ be unitary
transformations such that $U_2U_1=V_2V_1$ but $U_1\neq V_1$. Let
\[
P= U_1[q].c!0.U_2[q].\nil,\ \ \ \ Q= V_1[q].c!0.V_2[q].\nil.
\]
It is easy to check that $P\approx Q$ but $P\backslash
\{c\}\not\approx Q\backslash \{c\}$.

Now we turn to the congruence property of weak probabilistic
bisimilarity under the parallel combinator. First, we have some
lemmas.

\begin{lem}\label{lem:lem4.6}
For any configuration $<P;\bar{q}=\rho>$ and any super-operator $\e$
acting on $\h_{\bar{q}-qv(P)}$, we have
\begin{enumerate}
\item
$<P;\bar{q}=\rho>\rto{\qc
c?r:\sigma}<P';r,\bar{q}=\sigma\otimes\rho>\mbox{ if and only if }
<P;\bar{q}=\e(\rho)>\rto{\qc
c?r:\sigma}<P';r,\bar{q}=\sigma\otimes\e(\rho)>,$
\item
$<P;\bar{q}=\rho>\rto{\alpha}\boxplus{p_i}
\bullet<P_i;\bar{q}=\rho_i>\mbox{ if and only if }
<P;\bar{q}=\e(\rho)>\rto{\alpha}\boxplus{p_i}
\bullet<P_i;\bar{q}=\e(\rho_i)>$, where $\alpha$ is not of the form
$c?r:\sigma$.
\end{enumerate}
\end{lem}
{\it Proof.} (1) is obvious. For (2), we need only to prove the case
where $\alpha=\tau$ and the transition is due to a measurement. In
this case, if $<P;\bar{q}=\rho>\rto{\alpha}\boxplus{p_i}
\bullet<P_i;\bar{q}=\rho_i>$, then $\rho_i=P_{i,\bar{r}}\rho
P_{i,\bar{r}}/p_i$ for some projector $P_{i,\bar{r}}$ and
$p_i=\tr(P_{i,\bar{r}}\rho)$, where $\bar{r}\subseteq qv(P)$. So we
have $$<P;\bar{q}=\e(\rho)>\rto{\alpha}\boxplus{q_i}\bullet
<P_i;\bar{q}=P_{i,\bar{r}}\e(\rho) P_{i,\bar{r}}/q_i>$$ where
$q_i=\tr(P_{i,\bar{r}}\e(\rho))$. Notice that $\e$ is acting on
$\h_{\bar{q}-qv(P)}$ and $\bar{r}\subseteq qv(P)$. We deduce that
$$q_i=\tr(P_{i,\bar{r}}\e(\rho))=\tr\e(P_{i,\bar{r}}\rho P_{i,\bar{r}})=\tr(P_{i,\bar{r}}\rho)=p_i$$
and $P_{i,\bar{r}}\e(\rho) P_{i,\bar{r}}/q_i=\e(P_{i,\bar{r}}\rho
P_{i,\bar{r}}/q_i)$. That completes the proof of the necessity part.
The proof of the sufficiency part is similar. \hfill $\square$

\begin{lem}\label{lem:lem4.4}
If $<P;\bar{q}=\rho>\approx <Q;\bar{q}'=\rho'>$, then
$\bar{q}=\bar{q}'$, and $\tr_{\bar{r}}\rho=\tr_{\bar{r}}\rho'$ where
$\bar{r}=qv(P)\cup qv(Q)$.
\end{lem}
{\it Proof.} Suppose $\mathcal{G}_1$ and $\mathcal{G}_2$ are the
transition graphs of $<P;\bar{q}=\rho>$ and $<Q;\bar{q}'=\rho'>$,
respectively. Take a leaf $<P';C'>$ (so $<P';C'>\nrto{}$) of
$\mathcal{G}_1$ such that there exists a directed path from
$<P;\bar{q}=\rho>$ to $<P';C'>$ along which none of the actions has
the form $\qc c?q$. Intuitively, this path denotes an execution
where any quantum input action is realized by inputting a new qubit
from outside the context. As a result, the quantum system in
$\bar{q}-qv(P)$ is kept untouched in this path.

From the assumption that $<P;\bar{q}=\rho>\approx
<Q;\bar{q}'=\rho'>$, we can find a leaf $<Q';D'>$ of $\mathcal{G}_2$
such that $<P';C'>\approx <Q';D'>$ (so $C'=D'$), and furthermore,
there exists a directed path from $<Q;\bar{q}'=\rho'>$ to $<Q';D'>$
which has the same observable actions as the path taken in
$\mathcal{G}_1$. Notice that the set of quantum variables in the
accompanied context cannot be changed by $\tau$ actions. We deduce
$\bar{q}=\bar{q}'$ from the fact that $C'=D'$. Furthermore, we can
show $\tr_{\bar{r}}\rho=\tr_{\bar{r}}\sigma$ since the quantum
systems outside $\bar{r}$ are untouched during these two execution
paths.
 \hfill $\square$

\begin{lem}\label{lem:lem5.6}
Suppose $<P;\bar{q}=\rho>\approx <Q;\bar{q}=\rho'>$, $r\not\in
\bar{q}$ and $\sigma\in \dh$. Then
\begin{enumerate}
\item
$<P;r,\bar{q}=\sigma\otimes \rho>\approx <Q;r,\bar{q}=\sigma\otimes
\rho'>$.
\item If $P$ and $Q$ are free of quantum input and $\e$
is a super-operator acting on \\ $\h_{\bar{q}-qv(P)-qv(Q)}$, then
$<P;r,\bar{q}=\sigma\otimes \e(\rho)>\approx
<Q;r,\bar{q}=\sigma\otimes \e(\rho')>$.
\end{enumerate}
\end{lem}
{\it Proof.} We only prove (1). The proof of (2) is simpler since
$P$ and $Q$ are free of quantum input and as a result, the
super-operator $\e$ commutes with the quantum operations performed
by $P$ and $Q$. Let
\begin{eqnarray} \r'&=&\{(<P;r,\bar{q}=\sigma\otimes
\rho>, <Q;r,\bar{q}=\sigma\otimes \rho'>)\ \ |\
<P;\bar{q}=\rho>\approx <Q;\bar{q}=\rho'>, \nonumber\\
&&\hspace{2em}r\not\in \bar{q},\mbox{
and }\sigma\in \dh\}. 
\end{eqnarray}
We prove in the following that $\r=(\r'\cup \approx)^*$ is a weak
probabilistic bisimulation.

Suppose $(\c,\d)\in \r$. We may assume further that
$\c=<P;r,\bar{q}=\sigma\otimes \rho>$ and $\d=
<Q;r,\bar{q}=\sigma\otimes \rho'>$ for some $<P;\bar{q}=\rho>\approx
<Q;\bar{q}=\rho'>$, $r\not\in \bar{q}$, and $\sigma\in \dh$.

{\renewcommand{\theenumi}{(\roman{enumi})}
\begin{enumerate}

\item If $<P;r,\bar{q}=\sigma\otimes \rho>\rto{\alpha}\mu$, there are two cases to consider.
\begin{enumerate}
\item $\alpha=\qc c?r$ for some $\qc c\in qChan$. Then $
\mu=<P';r,\bar{q}=\sigma\otimes \rho>$ for some $P'$. By
\textbf{Q-Inp1} rule, we have $<P;\bar{q}=\rho>\rto{\qc
c?r:\sigma}\mu$. Now from the assumption $<P;\bar{q}=\rho>\approx
<Q;\bar{q}=\rho'>$, there exists a transition
$<Q;\bar{q}=\rho'>\Rto{\qc c?r:\sigma}\nu$ such that
$\mu\equiv_\approx \nu$. Thus it holds $<Q;r,\bar{q}=\sigma\otimes
\rho'>\Rto{\qc c?r}\nu$, and $\mu\equiv_\r \nu$ from the fact that
$\approx\subseteq \r$.

\item  $\alpha\not =\qc c?r$ for any $\qc c\in qChan$. Then we have
$<P;\bar{q}=\rho>\rto{\alpha}\mu_1=\boxplus p_i\bullet
<P_i;\bar{q}'=\rho_i>$ such that $r\not\in \bar{q}'$ and
$\mu=\boxplus p_i\bullet <P_i;r,\bar{q}'=\sigma\otimes \rho_i>$.
From the assumption $<P;\bar{q}=\rho>\approx <Q;\bar{q}=\rho'>$,
there exists a transition
$<Q;\bar{q}=\rho'>\Rto{\widehat{\alpha}}\nu_1=\boxplus q_j\bullet
<Q_j;\bar{q}'=\rho_j'>$ such that $\mu_1\equiv_\approx \nu_1$. So we
have $$<Q;r,\bar{q}=\sigma\otimes
\rho'>\Rto{\widehat{\alpha}}\nu=\boxplus q_j\bullet
<Q_j;r,\bar{q}'=\sigma\otimes \rho_j'>,$$ and $\mu\equiv_\r \nu$
from $\mu_1\equiv_\approx \nu_1$ and the definition of $\r$.
\end{enumerate}
\item  If $<P;r,\bar{q}=\sigma\otimes \rho>\nrto{}$ and $<Q;r,\bar{q}=\sigma\otimes \rho'>\nrto{}$,
then we have  $<P;\bar{q}= \rho>\nrto{}$ and $<Q;\bar{q}=
\rho'>\nrto{}$. Hence $\rho=\rho'$ from the assumption that
$<P;\bar{q}= \rho>\approx<Q;\bar{q}= \rho'>$, and then
$\sigma\otimes\rho =\sigma\otimes \rho'$.
\end{enumerate}
} From (i) and (ii) we know that $\r$ is a weak probabilistic
bisimulation on $Con$. That completes the proof of (1). \hfill
$\square$

\vspace{1em}

From the above lemmas, we are now ready to prove that weak
probabilistic bisimilarity is preserved by the parallel combinator
in two special cases, as the following two theorems state.
\begin{thm}\label{thm:noqinput}
If $P\approx Q$, and $P$ and $Q$ are free of quantum input, then
$P\| R\approx Q\| R$.
\end{thm}
{\it Proof.} Let
\begin{eqnarray*}
\r'&=&\{(<P\|R;\bar{q}=\e(\rho)>,<Q\|R;\bar{q}=\e(\rho')>)\ \ |\
<P;\bar{q}=\rho>\approx <Q;\bar{q}=\rho'>,\\&&\hspace{2em} \mbox{$P$
and $Q$ are free of quantum input,} \\
&& \hspace{3em}\mbox{and $\e$ is a super-operator on }
\h_{\bar{q}-qv(P)-qv(Q)}\}.
\end{eqnarray*}
We prove in the following that $\r=(\r'\cup \approx)^*$ is a weak
probabilistic bisimulation. Let
$(<P\|R;\bar{q}=\e(\rho)>,<Q\|R;\bar{q}=\e(\rho')>)\in \r'$.

{\renewcommand{\theenumi}{(\roman{enumi})}
\begin{enumerate}
\item Suppose $<P\|R;\bar{q}=\e(\rho)>\rto{\alpha}\mu$. Since
$P$ is free of quantum input, we have four cases to consider.
\begin{enumerate}
\item
There exists a transition
$<P;\bar{q}=\rho>\rto{\alpha}\mu_1=\boxplus{p_i}
\bullet<P_i;\bar{q}=\rho_i>$ where $qv(P_i)\subseteq qv(P)$ for each
$i$, and
$$\mu= \boxplus{p_i} \bullet<P_i\| R;\bar{q}=\e(\rho_i)>.$$ Here we
have used Lemma \ref{lem:lem4.6} (2). From the assumption
$<P;\bar{q}=\rho>\approx <Q;\bar{q}=\rho'>$, it holds that
$<Q;\bar{q}=\rho'>\Rto{{\widehat{\alpha}}}\nu_1=\boxplus{q_j}
\bullet<Q_j;\bar{q}=\rho_j'>$ and $\mu_1\equiv_\approx\nu_1$. Using
Lemma \ref{lem:lem4.6} (2) again, we derive
$$<Q\|R;\bar{q}=\e(\rho')>\Rto{\widehat{\alpha}}\nu=\boxplus{q_j}
\bullet<Q_j\| R;\bar{q}=\e(\rho_j')>,$$ and
 $\mu\equiv_{\r}\nu$ from the fact that $qv(P_i)\subseteq qv(P)$ for each
$i$, $\mu_1\equiv_\approx \nu_1$, and the definition of $\r$.

\item  There exists a transition
$<R;\bar{q}=\e(\rho)>\rto{\qc c?r:\sigma}
<R';r,\bar{q}=\sigma\otimes \e(\rho)>$ for some $\qc c\in qChan$,
$r\not\in \bar{q}$, $\sigma\in \dh$, and
$\mu=<P\|R';r,\bar{q}=\sigma\otimes \e(\rho)>$. Then from
\textbf{Q-Inp1} and \textbf{Inp-Int} rules, we have
$<R;\bar{q}=\e(\rho')>\rto{\qc c?r:\sigma}
<R';r,\bar{q}=\sigma\otimes \e(\rho')>$ and so
$$<Q\|R;\bar{q}=\e(\rho')>\rto{\qc
c?r:\sigma}<Q\|R';r,\bar{q}=\sigma\otimes \e(\rho')>.$$ Furthermore,
we can prove $(<P\|R';r,\bar{q}=\sigma\otimes \e(\rho)>,
<Q\|R';r,\bar{q}=\sigma\otimes \e(\rho')>)\in\r$ by Lemma
\ref{lem:lem5.6} (2).

\item There exists a transition $<R;\bar{q}=\e(\rho)>\rto{\alpha}\boxplus
{p_i}\bullet<R_i;\bar{q}=\e_i(\e(\rho))>$ where $\alpha$ is not of
the form $\qc c?r:\sigma$, $\e_i$ is a super-operator on
$\mathcal{L}(\h_{qv(R)})$, and $\mu=\boxplus
{p_i}\bullet<P\|R_i;\bar{q}=\e_i(\e(\rho))>$. Here we have used
Lemma \ref{lem:superoperator}. Then from Lemma \ref{lem:lem4.4}, we
derive $<R;\bar{q}=\e(\rho')>\rto{\alpha}\boxplus
{p_i}\bullet<R_i;\bar{q}=\e_i(\e(\rho'))>$. Thus
$$<Q\|R;\bar{q}=\e(\rho')>\rto{\alpha}\nu=\boxplus
{p_i}\bullet<Q\|R_i;\bar{q}=\e_i(\e(\rho'))>.$$ Notice that for any
$i$, we have
$(<P\|R_i;\bar{q}=\e_i(\e(\rho))>,<Q\|R_i;\bar{q}=\e_i(\e(\rho'))>)\in\r$
since the composite map $\e^i\circ\e$ is also a super-operator
acting on $\h_{\bar{q}-qv(P)-qv(Q)}$. Then it follows that
$\mu\equiv_\r \nu$.

\item $\alpha=\tau$, and the action is caused by a
communication between $P$ and $R$. Without loss of any generality,
we assume that
$$<P;\bar{q}=\e(\rho)>\rto{c?v}{<P';\bar{q}=\e(\rho)>},\ \ \  <R;\bar{q}=\e(\rho)>\rto{c!v}{<R';\bar{q}=\e(\rho)>}$$
where $qv(P')=qv(P)$ and $\mu={<P'\|R';\bar{q}=\e(\rho)>}$. Then
$<P;\bar{q}=\rho>\rto{c?v}{<P';\bar{q}=\rho>}$, and from the
assumption $<P;\bar{q}=\rho>\approx <Q;\bar{q}=\rho'>$, we derive
that
$$<Q;\bar{q}=\rho'>\Rto{c?v}\boxplus{p_i}\bullet<Q_i;\bar{q}=\rho_i'>,$$
and for any $i$, $<P';\bar{q}=\rho>\approx <Q_i;\bar{q}=\rho_i'>.
$ Notice that from
$<R;\bar{q}=\e(\rho)>\rto{c!v}<R';\bar{q}=\e(\rho)>$ we can deduce
that $<R;C>\rto{c!v}<R';C>$ for any context $C$. Thus
$$<Q\|R;\bar{q}=\rho'>\Rto{{\tau}}\nu=\boxplus{p_i}\bullet<Q_i\|R';\bar{q}=\rho_i'>$$
by using \textbf{C-Com} rule. Furthermore, we have $\mu\equiv_\r\nu$
since $(<P'\|R';\bar{q}=\e(\rho)>, <Q_i\|R';\bar{q}=\e(\rho_i')>)\in
\r $ for each $i$, which in turn can be be proved by the facts that
$qv(P')=qv(P)$ and $<P';\bar{q}=\rho>\approx <Q_i;\bar{q}=\rho_i'>$.
\end{enumerate}
\item If $<P\|R;\bar{q}=\e(\rho)>\nrto{}$ and $<Q\|R;\bar{q}=\e(\rho')>\nrto{}$, then we have
$<P;\bar{q}=\rho>\nrto{}$ and $<Q;\bar{q}=\rho'>\nrto{}$ . Hence
$\rho=\rho'$ from the assumption $<P;\bar{q}=\rho>\approx
<Q;\bar{q}=\rho'>$. So we derive $\e(\rho)=\e(\rho')$.
\end{enumerate}
} From (i) and (ii) we know that $\r$ is a weak probabilistic
bisimulation on $Con$. For any quantum context $\bar{q}=\rho$, by
$P\approx Q$ we have $<P;\bar{q}=\rho>\approx <Q;\bar{q}=\rho>$ and
then $(<P\| R;\bar{q}=\rho>, <Q\| R;\bar{q}=\rho>)\in \r$ since the
identity transformation is also a super-operator on
$\h_{\bar{q}-qv(P)-qv(Q)}$. Then it follows that $<P\|
R;\bar{q}=\rho>\approx <Q\| R;\bar{q}=\rho>$, and so $P\|R\approx
Q\|R$ from the arbitrariness of the context. \hfill $\square$

\vspace{1em}

The constraint that $P$ and $Q$ are free of quantum input is vital
for the proof of this theorem: it guarantees that for any derivative
$<P';C>$ (node in the transition graph) of $<P;\bar{q}=\rho>$,
$qv(P')\subseteq qv(P)$, and then, any super-operator $\e$ acting on
$\h_{\bar{q}-qv(P)}$ is also a super-operator acting on
$\h_{\bar{q}-qv(P')}$. As a result, any quantum unitary
transformation or measurement performed by $<P';C>$ commutes with
$\e$. When $P$ and $Q$ are not free of quantum input, an example
(see Example \ref{exam:5.1} below) will be presented to show why the
proof technique used in this theorem fails.

Although we only consider in Theorem \ref{thm:noqinput} a special
case where neither $P$ nor $Q$ will ever have the power to input a
qubit, this case covers an important scenario called LOCC (local
operations and classical communication) in quantum information
field. When communicating parties are spatially separated, they are
usually restricted to performing local (quantum) operations on their
own subsystems and transmitting classical information (say, the
outcomes of measurements) to coordinate the local operations. This
restriction is partially due to technological consideration:
noiseless long-distance quantum communication is often very
difficult to realize. LOCC restriction is also widely required in
the study of quantum entanglement \cite{Mi99, NC00}.

\begin{thm}\label{thm:nouandm}
If $P\approx Q$, then $P\| R\approx Q\| R$ provided that $R$ is free
of unitary transformation and quantum measurement.
\end{thm}
{\it Proof.} Let
\begin{eqnarray*}
\r'&=&\{(<P\|R;C>,<Q\|R;D>)\ \ |\ <P;C>\approx <Q;D>, \\
&&\hspace{2em} \mbox{$R$ is free of unitary transformation and
quantum measurement}\}.
\end{eqnarray*}
We prove in the following that $\r=(\r'\cup \approx)^*$ is a weak
probabilistic bisimulation. Suppose $(<P\|R;C>,<Q\|R;D>)\in \r'$.

{\renewcommand{\theenumi}{(\roman{enumi})}
\begin{enumerate}

\item If $<P\|R;C>\rto{\alpha}\mu$, there are four cases to consider.
\begin{enumerate}
\item
There exists a transition $<P;C>\rto{\alpha}\mu_1=\boxplus{p_i}
\bullet<P_i;C_i>$ and $\mu= \boxplus{p_i} \bullet<P_i\| R;C_i>$. By
the assumption that $<P;C>\approx <Q;D>$, we have
$<Q;D>\Rto{{\widehat{\alpha}}}\nu_1=\boxplus{q_j} \bullet<Q_j;D_j>$
such that $\mu_1\equiv_\approx\nu_1$. So it holds
$$<Q\|R;D>\Rto{\widehat{\alpha}}\nu=\boxplus{q_j} \bullet<Q_j\|
R;D_j>.$$ Furthermore, we can prove $\mu\equiv_{\r}\nu$ from
$\mu_1\equiv_\approx \nu_1$ and the definition of $\r$.

\item  There exists a transition
$<R;\bar{q}=\rho>\rto{\qc c?r:\sigma} <R';r,\bar{q}=\sigma\otimes
\rho>$ for some $\qc c\in qChan$, $r\not\in \bar{q}$, $\sigma\in
\dh$, and $\mu=<P\|R';r,\bar{q}=\sigma\otimes \rho>$. Here we assume
that $C$ and $D$ are of the forms $\bar{q}=\rho$ and
$\bar{q}=\rho'$, respectively. Then from \textbf{Q-Inp1} and
\textbf{Inp-Int} rules, we have
$$<Q\|R;\bar{q}=\rho'>\rto{\qc c?r:\sigma}<Q\|R';r,\bar{q}=\sigma\otimes
\rho'>,$$ and $(<P\|R';r,\bar{q}=\sigma\otimes \rho>,
<Q\|R';r,\bar{q}=\sigma\otimes \rho'>)\in\r$ from Lemma
\ref{lem:lem5.6} (1) and the fact that $R'$ is also free of unitary
transformation and quantum measurement.

\item There exists a transition $<R;C>\rto{\alpha}<R';C>$ where $\alpha$ is not
of the form $\qc c?r:\sigma$, and $\mu=<P\|R';C>$. Here we have used
the assumption that $R$ is free of unitary transformation and
quantum measurement. Then it holds that $<R;D>\rto{\alpha}<R';D>$
and then
$$<Q\|R;D>\rto{\alpha}<Q\|R';D>.$$ Furthermore, we have
$(<P\|R';C>,<Q\|R';D>)\in \r$ by the definition of $\r$.

\item  $\alpha=\tau$, and the action is caused by a (classical or
quantum) communication between $P$ and $R$. We assume that
$$<P;C>\rto{\qc c?r}{<P';C>},\ \ \  <R;C>\rto{\qc
c!r}{<R';C>}$$ and $\mu={<P'\|R';C>}$. Other cases are similar. From
the assumption that $<P;C>\approx <Q;D>$, we have
$$<Q;D>\Rto{\qc c?r}\boxplus{p_i}\bullet<Q_i;D_i>
\mbox{ and for any $i$ }, <P';C>\approx <Q_i;D_i>.
$$
 Notice that from $<R;C>\rto{\qc
c!r}{<R';C>}$ we can deduce that  $<R;G>\rto{\qc c!r}{<R';G>}$ for
any context $G$ involving the qubit $r$. Thus from \textbf{Q-Com}
rule,
\[
<Q\|R;D>\Rto{{\tau}}\nu=\boxplus{p_i}\bullet<Q_i\|R';D_i>.
\]
In order to show $\mu\equiv_\r\nu$, we need only to prove that for
any $i$, $(<P'\|R';C>, <Q_i\|R';D_i>)\in \r$, which is direct from
the fact that $<P';C>\approx <Q_i;D_i>$.
\end{enumerate}
\item  If $<P\|R;C>\nrto{}$ and $<Q\|R;D>\nrto{}$, then we have $<P;C>\nrto{}$ and $<Q;D>\nrto{}$. Hence
$C=D$ from the assumption $<P;C>\approx<Q;D>$.
\end{enumerate}
} From (i) and (ii) we know that $\r$ is a weak probabilistic
bisimulation on $Con$. So by $P\approx Q$, we can deduce that
$<P;C>\approx <Q;C>$ for any context $C$. Then
$(<P\|R;C>,<Q\|R;C>)\in \r$ and hence $<P\|R;C>\approx <Q\|R;C>$.
Finally, we derive $P\|R\approx Q\|R$ by the arbitrariness of $C$.
\hfill $\square$

\vspace{1em}

As we know, the standard technique in classical process algebra for
proving that bisimilarity is preserved by static combinators such as
relabeling, restriction, and parallel combinators is to construct a
relation consisting of pairs of configurations having the considered
static structure, and prove that it is a bisimulation. This
technique is also used in the proofs of Theorems
\ref{lem:wprepreserve}, \ref{lem:wrelpreserve}, and
\ref{thm:noqinput}. It will fail, however, to prove the congruence
property under parallel combinator when general quantum processes
are considered. The following example illustrates how entanglement
between different quantum systems and the non-commutativity of
quantum operations make the technique fail. Particularly, we will
construct quantum processes $P$, $Q$, $R$, and context $C$, such
that $<P;C>\approx <Q;C>$ but $<P\| R; C>\not\approx <Q\| R; C>$.
\begin{exmp}\label{exam:5.1}\rm
Let $M_{0,1}$, $\sigma_0$, $\sigma_1$, and $|+\>$ be given as in
Section 2 and Example \ref{exam:3.2}. Suppose $P=\qc
c?q.M_{0,1}[q;x].\nil$, and
$$Q=\qc c?q.(M_{0,1}[q;x].\sigma_x[q].\nil +
M_{0,1}[q;x].\sigma_{1-x}[q].\nil)$$ is the process which inputs a
qubit and then nondeterministically sets it to $|0\>$ or $|1\>$. Let
$\c=<P;q=|+\>\<+|>$ and $\d=<Q;q=|+\>\<+|>$. Then the transition
graphs of $\c$ and $\d$ can be depicted respectively as Fig.~\ref{fig:temp}
\begin{figure}[t]
 \centering \includegraphics[width=0.9\textwidth]{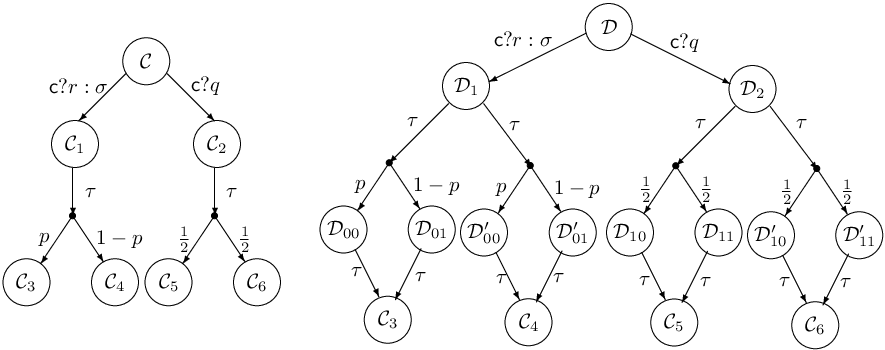} 
\caption{The transition graphs for $\c$ and $\d$ in Example~\ref{exam:5.1}.\label{fig:temp}}
\end{figure} 
where $p= \<0|\sigma|0\>$ and
\[\begin{array}{ll}
 \c_1 = <M_{0,1}[r;x].\nil; r,q=\sigma\otimes|+\>\<+|>,&\c_2 = <M_{0,1}[q;x].\nil;
 q=|+\>\<+|>,\\ 
  \c_3 = <\nil; r,q=|0\>\<0|\otimes|+\>\<+|>,\ \ &    \c_4 = <\nil;
  r,q=|1\>\<1|\otimes|+\>\<+|>,\\ 
\c_5 = <\nil; q=|0\>\<0|>,\ \ &\c_6 = <\nil; q=|1\>\<1|>,\\ 
 \d_1 = <Q'[r/q]; r,q=\sigma\otimes|+\>\<+|>,\ \
&\d_2 = <Q';
 q=|+\>\<+|>,\\ 
  \d_{0i} = <\sigma_i[r].\nil; r,q=|i\>\<i|\otimes|+\>\<+|>,\ \ &      \d_{0i}' = <\sigma_{1-i}[r].\nil; r,q=|i\>\<i|\otimes|+\>\<+|>,\\ 
  \d_{1i} = <\sigma_i[q].\nil; q=|i\>\<i|>,\ \ &      \d_{1i}' = <\sigma_{1-i}[q].\nil;
  q=|i\>\<i|>,
  \end{array}
\]
and $$Q'=  M_{0,1}[q;x].\sigma_x[q].\nil +
M_{0,1}[q;x].\sigma_{1-x}[q].\nil.$$ Take
$$\r=\{(\c,\d),(\c_1,\d_1),(\c_2,\d_2),(\c_3,\d_{0i}),(\c_4,\d_{0i}'),(\c_5,\d_{1i}),(\c_6,\d_{1i}') : i=0,1\}.$$
It is easy to check that $\r$ is indeed a weak probabilistic
bisimulation. Thus $\c\approx \d$.

Now let $R=\qc c?r.CNOT[q,r].\qc c!q.\nil.$ Then we have
$$<P\| R;q=|+\>\<+|>\not\approx <Q\| R;q=|+\>\<+|>$$ because $<P\| R;q=|+\>\<+|>$ has a transition sequence
\begin{eqnarray*} <P\| R;q=|+\>\<+|>
&\rto{\qc c?r:|0\>\<0|}& <P\| (CNOT[q,r].\qc
c!q.\nil); r,q=[|0\>|+\>]>\\
&\rto{\tau}&<P\| \qc
c!q.\nil; r,q=[\frac{1}{\sqrt{2}}(|00\>+|11\>)]>\\
&\rto{\tau}&<M_{0,1}[q;x].\nil\| \nil; r,q=[\frac{1}{\sqrt{2}}(|00\>+|11\>)]>\\
&\rto{\tau}&\frac{1}{2}\bullet<\nil\| \nil;
r,q=[|00\>]>\\
&&\ \  \ \boxplus\frac{1}{2}\bullet<\nil\| \nil; r,q=[|11\>]>
\end{eqnarray*}
while the only form of combined weak $\qc c?r:|0\>\<0|$-transitions
of $<Q\| R;q=|+\>\<+|>$ is
\begin{eqnarray*}<Q\| R;q=[|+\>]> &\Rto{\qc c?r:|0\>\<0|}&s\bullet<\nil\| \nil;
r,q=[|00\>]>\\
&&\ \  \boxplus(1-s)\bullet<\nil\| \nil; r,q=[|01\>]>
\end{eqnarray*}
where $s\in [0,1]$.  \hfill $\square$
\end{exmp}
\subsection{Equality relation between quantum processes}

As in classical process algebra, $\approx$ is not preserved by
summation combinator `+'. To deal with it, we introduce the notion
of equality between quantum processes.

\begin{defn}\label{def:equality}
Two configurations $\c$ and $\d$ are said to be equal, denoted by
$\c\simeq\d$, if for any $\alpha\in Act$,
\begin{enumerate}
\item
whenever $\c \rto{\alpha} \mu$ then there exists $\nu$ such that $\d
\Rto{\alpha} \nu$ and $\mu\equiv_\approx\nu$,
\item
whenever $\d \rto{\alpha} \nu$ then there exists $\mu$ such that $\c
\Rto{\alpha} \mu$ and $\mu\equiv_\approx\nu$,
\item
if $\c\nrto{}$ and $\d\nrto{}$, then $Contex(\c)=Contex(\d)$.
\end{enumerate}
\end{defn}

The only difference between the definitions of $\approx$ and
$\simeq$ is that in the latter $\d \Rto{\widehat{\alpha}} \nu$ is
replaced by $\d \Rto{\alpha} \nu$, $i.e.$, the matching action for a
$\tau$-move has to be a real $\tau$-move.

Furthermore, we lift the definition of equality to quantum processes
as follows. For $P,Q\in qProc$, and $\bar{x}$ is the set of free
classical variables contained in $P$ and $Q$, $P\simeq Q$ if
$P[\bar{v}/\bar{x}]\simeq Q[\bar{v}/\bar{x}]$ for any indexed set
$\bar{v}$ of values.

The following properties are direct from definition. So we omit the
proofs here.

\begin{thm}
$P\sim Q$ implies $P\simeq Q$, and $P\simeq Q$ implies $P\approx Q$.
\end{thm}

\begin{thm}
If $P\approx Q$ then $a.P\simeq a.Q$ for any $a\in \{c?x,c!e,\qc
c?q,\qc c!q,U[\bar{q}],M[\bar{q};x]\}$;
\end{thm}

\begin{thm}\label{lem:summation}
For any $P,Q\in qProc$, $P\simeq Q$ if and only if $P+R\approx Q+R$
for all $R\in qProc$.
\end{thm}

Finally, a congruence property similar to Theorem
\ref{lem:sprepreserve} is also satisfied by the quality relation.

\begin{thm} If $P\simeq Q$ then
\begin{enumerate}
\item
 $a.P\simeq a.Q$, for any $a\in \{c?x,c!e,\qc c?q,\qc
c!q,U[\bar{q}],M[\bar{q};x]\}$,

\item $P+R\simeq Q+R$, for any $R\in qProc$,

\item $P\|R\simeq Q\|R$, provided that $R$ is free of unitary
transformation and measurement, or $P$ and $Q$ are free of quantum
input,

\item $P[f]\simeq Q[f]$, for any relabeling function $f$,

\item $\iif\ b\ \then\ P\simeq \iif\ b\ \then\ Q$
 for any boolean expression $b$.
\end{enumerate}
\end{thm}
{\it Proof.} (2) is direct from Theorem \ref{lem:summation}. Others
are similar to the proofs of corresponding results for $\approx$.
\hfill $\square$
\section{Conclusions and further work}

In this paper, we propose a framework qCCS to model and reason about
the behaviors of quantum concurrent systems. This framework is a
natural quantum extension of classical value-passing CCS. To make
qCCS consistent with the laws of quantum mechanics, some syntactical
restrictions on valid quantum processes are introduced. The
operational semantics of qCCS is given in terms of probabilistic
labeled transition system. This semantics has many different
features compared with the proposals in literature in describing
input and output of quantum systems which are correlated with other
systems. We make the design decision of keeping the probability
information resulting from quantum measurements instead of resolving
probabilistic choice in each intermediate step as is done in
\cite{JL04} and \cite{GN05}. Based on this operational semantics, we
define the notions of strong (weak) probabilistic bisimulation and
equality between quantum processes and examine some properties such
as congruence of them.

The congruence property we proved in this paper is, however, a weak
one in which bisimilarity is preserved by the parallel combinator
when some constraints are put on paralleled processes. New
techniques must be invented when general processes are considered,
since we have presented an example to show why standard proof
techniques do not work because of the entanglement between quantum
systems and the non-commutativity of quantum operations. A potential
way to tackle this problem, motivated by Theorem
\ref{lem:summation}, is to define a new relation, say $\sim'$,
between quantum processes such that $P\sim' Q$ if and only if for
any $R$, $P\| R \sim Q\|R$. Obviously we have $\sim'\subseteq \sim$,
and $\sim'$ is also an equivalence relation. Furthermore, we can
show that this relation is preserved by all combinators defined in
this paper except for restriction. So the problem of whether strong
probabilistic bisimilarity is preserved by the parallel combinator
is equivalent to the problem of whether or not $\sim'=\sim$.

Another direction along this line is to give up the notion of
bisimulation and instead search for other coarser order relations
among quantum processes which are preserved by the combinators
defined in this paper. For example, we can drop the symmetry of
bisimulation and instead define a notion of simulation which relates
processes $P$ and $Q$ if for any context $C$, each action of $<P;C>$
can be simulated by a (combined) action of $<Q;C>$, and the resulted
configurations also satisfy this order relation.

Recursive definitions are very useful in modeling infinite behavior
of processes. Furthermore, uniqueness of solutions of recursion
equations provides a powerful tool for reasoning about the
correctness of implementations with respect to specifications.
However, there are some technical difficulties in introducing
recursive constructs into qCCS. For example, if we allow the process
 defined by
\begin{equation}\label{eq:recursion}
A:=\qc c!q.A \end{equation} to be valid, then problems will occur
when we attempt to assign free quantum variables to $A$: on one
hand, from Definition \ref{def:qProc} (5), to make $\qc c!q.A$
meaningful we must have $q\not\in qv(A)$; on the other hand, also
from Definition \ref{def:qProc} (5), we know $q\in qv(\qc c!q.A)$.
This is a contradiction because we will naturally require $qv(\qc
c!q.A)\subseteq qv(A)$ in definition equation (\ref{eq:recursion}).
However, the difficulty does not exist in the following recursively
defined quantum process
\begin{equation}
A:=\qc c?q.U[q].\qc c!q.A \end{equation} which consequently inputs a
qubit through quantum channel $\qc c$, applies a predefined unitary
transformation $U$ on it, and outputs it through $\qc c$. Here we
can freely let $qv(A)=\emptyset$.

In order to provide some useful mathematical tools for describing
approximate correctness and evolution of concurrent systems, one of
the authors has tried to develop topology in process algebras
\cite{Yi01}. In particular, he and Wirsing \cite{YW00} introduced
the notions of $\lambda$-bisimulation and approximate bisimulation
in CCS equipped with a metric on its set of action names, and
further applied them to probabilistic processes \cite{Yi02b}. To
extend these notions to the quantum setting is a direction worthy of
future investigation.

\section*{Acknowledgement}
We thank the referees for their helpful comments and suggestions,
which improved the presentation and the quality of this paper.

The authors thank the colleagues in the Quantum Computation and
Quantum Information Research Group for useful discussion. This work
was partially supported by the FANEDD under Grant No.~200755, the
863 Project under Grant No.~2006AA01Z102, and the Natural Science
Foundation of China (Grant Nos.~60503001, 60621062, and 60433050).
Y. Feng was also partly supported by Tsinghua Basic Research
Foundation under Grant No. 052220204.

\bibliographystyle{plain}

\bibliography{qProcess}

\begin{thebibliography}{10}

\bibitem{Be92}
C.~H. Bennett.
\newblock Quantum cryptography using any two nonorthogonal states.
\newblock {\em Physical Review Letters}, 68:3121, 1992.

\bibitem{BB84}
C.~H. Bennett and G.~Brassard.
\newblock Quantum cryptography: Public-key distribution and coin tossing.
\newblock In {\em Proceedings of the IEEE International Conference on Computer,
  Systems and Signal Processing}, pages 175--179, Bangalore, India, 1984.

\bibitem{BB93}
C.~H. Bennett, G.~Brassard, C.~Crepeau, R.~Jozsa, A.~Peres, and W.~Wootters.
\newblock Teleporting an unknown quantum state via dual classical and epr
  channels.
\newblock {\em Physical Review Letters}, 70:1895--1899, 1993.

\bibitem{BW92}
C.~H. Bennett and S.~J. Wiesner.
\newblock Communication via one- and two-particle operators on
  einstein-podolsky-rosen states.
\newblock {\em Physical Review Letters}, 69(20):2881--2884, 1992.

\bibitem{BCS03}
S.~Bettelli, T.~Calarco, and L.~Serafini.
\newblock Toward an architecture for quantum programming.
\newblock {\em European Physical Journal D}, 25(2):181--200, 2003.

\bibitem{Bu87}
P.~J. Bussey.
\newblock Communication and non-communication in einstein-rosen experiments.
\newblock {\em Physics Letters A}, 123:1--3, 1987.

\bibitem{Ek91}
A.~K. Ekert.
\newblock Quantum cryptography based on bell's theorem.
\newblock {\em Physical Review Letters}, 67:661, 1991.

\bibitem{Fe82}
R.~Feynman.
\newblock Simulating physics with computers.
\newblock {\em International Journal of Theoretical Physics}, 21:467--488,
  1982.

\bibitem{Ga06}
S.~J. Gay.
\newblock Quantum programming languages: survey and bibliography.
\newblock {\em Mathematical Structures in Computer Science}, 16(04):581--600,
  2006.

\bibitem{GN05}
S.~J. Gay and R.~Nagarajan.
\newblock Communicating quantum processes.
\newblock In J.~Palsberg and M.~Abadi, editors, {\em Proceedings of the 32nd
  ACM SIGPLAN-SIGACT Symposium on Principles of Programming Languages (POPL)},
  pages 145--157, 2005.

\bibitem{GW83}
G.~C. Ghirardi and T.~Weber.
\newblock Quantum mechanics and faster-than-light communication methodological
  considerations.
\newblock {\em Nuovo Cimento B}, 11(78 B):9--20, 1983.

\bibitem{Gr96}
L.~K. Grover.
\newblock A fast quantum mechanical algorithm for database search.
\newblock In {\em Proc. ACM STOC}, pages 212--219, 1996.

\bibitem{Gr97}
L.~K. Grover.
\newblock Quantum mechanics helps in searching for a needle in a haystack.
\newblock {\em Physical Review Letters}, 78(2):325, 1997.

\bibitem{He91}
M.~Hennessy.
\newblock A proof system for communicating processes with value-passing.
\newblock {\em Formal Aspects of Computer Science}, 3:346--366, 1991.

\bibitem{HI93}
M.~Hennessy and A.~Ing\'{o}lfsd\'{o}ttir.
\newblock A theory of communicating processes value-passing.
\newblock {\em Information and Computation}, 107(2):202--236, 1993.

\bibitem{JL04}
P.~Jorrand and M.~Lalire.
\newblock Toward a quantum process algebra.
\newblock In P.~Selinger, editor, {\em Proceedings of the 2nd International
  Workshop on Quantum Programming Languages, 2004}, page 111, 2004.

\bibitem{Kn96}
E.~H. Knill.
\newblock Conventions for quantum pseudocode.
\newblock {\em LANL report LAUR-96-2724}, 1996.

\bibitem{Kr83}
K.~Kraus.
\newblock {\em States, Effects and Operations: Fundamental Notions of Quantum
  Theory}.
\newblock Springer, Berlin, 1983.

\bibitem{La05}
Marie Lalire.
\newblock A probabilistic branching bisimulation for quantum processes.
\newblock 2005.
\newblock arXiv:quant-ph/0508116 v1 16 Aug 2005.

\bibitem{La06}
Marie Lalire.
\newblock Relations among quantum processes: Bisimilarity and congruence.
\newblock {\em Mathematical Structures in Computer Science}, 16(3):407--428,
  2006.

\bibitem{LS91}
K.~Larsen and A.~Skou.
\newblock Bisimulation through probabilistic testing.
\newblock {\em Information and Computation}, 94:456--471, 1991.

\bibitem{MP92}
R.~Milner, J.~Parrow, and D.~Walker.
\newblock A calculus of mobile processes, parts i and ii.
\newblock {\em Information and Computation}, 100:1--77, 1992.

\bibitem{Mi99}
M.~Nielsen.
\newblock Conditions for a class of entanglement transformations.
\newblock {\em Physical Review Letters}, 83:436--439, 1999.

\bibitem{NC00}
Michael Nielsen and Isaac Chuang.
\newblock {\em Quantum computation and quantum information}.
\newblock Cambridge university press, 2000.

\bibitem{Om98}
B.~\"{O}mer.
\newblock {\em A procedural formalism for quantum computing}.
\newblock Master thesis, Department of Theoretical Physics, Technical
  University of Vienna, 1998.
\newblock http://tph.tuwien.ac.at/oemer/qcl.html.

\bibitem{Om03}
B.~\"{O}mer.
\newblock {\em Structured Quantum Programming}.
\newblock PhD thesis, Department of Theoretical Physics, Technical University
  of Vienna, 2003.

\bibitem{PF04}
A.~Poppe, A.~Fedrizzi, T.~Lor¨unser, O.~Maurhardt, R.~Ursin, H.~R. B¨ohm,
  M.~Peev, M.~Suda, C.~Kurtsiefer, H.~Weinfurter, T.~Jennewein, and
  A.~Zeilinger.
\newblock Practical quantum key distribution with polarization entangled
  photons.
\newblock 2004.
\newblock arXiv:quant-ph/0404115.

\bibitem{SZ00}
J.~W. Sanders and P~Zuliani.
\newblock Quantum programming.
\newblock {\em Mathematics of Program Construction}, 1837:80--99, 2000.

\bibitem{SL94}
R.~Segala and N.~Lynch.
\newblock Probabilistic simulations for probabilitsic processes.
\newblock In {\em Proc. CONCUR'94, Theories of Concurrency Unification and
  Extension, Lecture Notes in Computer Science}, volume 836, pages 481--496,
  1994.

\bibitem{SL95}
R.~Segala and N.~Lynch.
\newblock Probabilistic simulations for probabilistic processes.
\newblock {\em Nordic Journal of Computing}, 2(2):250--273, 1995.

\bibitem{Se041}
P.~Selinger.
\newblock A brief survey of quantum programming languages.
\newblock {\em Functional and Logic Programming}, 2998:1--6, 2004.

\bibitem{Se04}
P.~Selinger.
\newblock Towards a quantum programming language.
\newblock {\em Mathematical Structures in Computer Science}, 14(4):527--586,
  2004.

\bibitem{Sh94}
Peter~W. Shor.
\newblock Algorithms for quantum computation: discrete log and factoring.
\newblock In {\em Proceedings of the 35th IEEE FOCS}, pages 124--134, 1994.

\bibitem{vN55}
J.~von Neumann.
\newblock {\em Mathematical Foundations of Quantum Mechanics}.
\newblock Princeton University Press, Princeton, NJ, 1955.

\bibitem{WZ82}
W.~K. Wootters and W.~H. Zurek.
\newblock A single quantum cannot be cloned.
\newblock {\em Nature}, 299(5886):802--803, 1982.

\bibitem{YW00}
M~.~S. Ying and M.~Wirsing.
\newblock Approximate bisimilarity.
\newblock In T.~Rus, editor, {\em Algebraic Methodology and Software
  Technology, 8th International Conference}, volume 1816 of {\em Lecture Notes
  in Computer Science}, pages 309--321, Iowa City, USA, 2000.

\bibitem{Yi01}
M.~S. Ying.
\newblock {\em Topology in Process Calculus: Approximate Correctness and
  Infinite Evolution of Concurrent Programs}.
\newblock Springer-Verlag New York, 2001.

\bibitem{Yi02b}
M.~S. Ying.
\newblock Additive models of probabilistic processes.
\newblock {\em Theoretical Computer Science}, 275:481--519, 2002.

\bibitem{Zu01}
P.~Zuliani.
\newblock {\em Quantum Programming}.
\newblock PhD thesis, Oxford University, 2001.

\bibitem{Zu05b}
P~Zuliani.
\newblock Quantum programming with mixed states.
\newblock In {\em Proceedings of the 3rd International Workshop on Quantum
  Programming Languages}, Chicago, 2005.

\end{thebibliography}

\end{document}